\documentclass[11pt]{article}
\usepackage{graphicx}
\usepackage{amsmath}
\usepackage{indentfirst}
\usepackage{amsfonts}
\usepackage{amssymb}
\usepackage{setspace}
\usepackage{booktabs}
\usepackage{cancel}
\usepackage{float}
\usepackage{bm,bbm}
\usepackage[T1]{fontenc}
\usepackage{textcomp}
\usepackage{listings}
\usepackage{kotex}
\usepackage{caption}
\usepackage{subcaption}
\usepackage{times}
\usepackage{hyperref}

\usepackage[colorinlistoftodos,prependcaption,textsize=tiny]{todonotes}
\usepackage[backend=bibtex,sorting=none,style=numeric-comp]{biblatex}
\usepackage[font=normal,format=plain,labelfont=bf]{caption}


\numberwithin{equation}{section}
\newcommand{\eref}[1]{(\ref{#1})} 
\newcommand{\fref}[1]{Fig.~\ref{#1}}
\newcommand{\sref}[1]{Section~\ref{#1}}
\newcommand{\bu}{\mathbf{u}}
\newcommand{\itn}{\textit{n}}

\begin{document}

\begin{center}
\textbf{\Large Numerical Modeling of n-Hexane Pyrolysis with an Optimized Kinetic Mechanism in a Hydrogen Plasma Reactor}

\bigskip

Subin Choi$^{1,2}$, Chanmi Jung$^{1}$, Dae Hoon Lee$^{1}$, Jeongan Choi$^{1,*}$, Jaekwang Kim$^{2,*}$\\
\bigskip
\small{
\textit{
$^1$Semiconductor Manufacturing Research Center, Korea Institute of Machinery and Materials,\\ Daejeon, Republic of Korea\\
$^2$Department of Mechanical and Design Engineering, Hongik University,\\ Sejong, Republic of Korea\\
}

\bigskip
sbchoi@kimm.re.kr, chanmi@kimm.re.kr,  dhlee@kimm.re.kr, \\jachoi@kimm.re.kr, jk12@hongik.ac.kr\\
$^*$ Corresponding author 
}
\end{center}

\bigskip

\begin{center}
\textbf{Abstract }\\
\bigskip
\begin{minipage}{0.85\textwidth}
The physicochemical mechanisms underlying the pyrolysis of \textit{n}-hexane in a high temperature Ar-H\textsubscript{2} environment were investigated for plasma pyrolysis process. 
An optimal chemical kinetics model was developed using the Reaction Mechanism Generator (RMG), an automated tool for constructing reaction mechanisms.
This model was validated through 0-D analyses, where simulation result were compared with existing kinetic models (LLNL, JetSurf) and experimental data from conventional \textit{n}-hexane pyrolysis. 
Subsequently, 1-D analysis were conducted to identify the optimal operational flow rate in plasma pyrolysis reactor, the results of which informed detailed three-dimensional (2-D) computational fluid dynamics (CFD) modeling of the plasma reactor.
The CFD simulations reveal that fluid mixing dynamics play a dominant role in determining the extent of conversion and product selectivity, highlighting the limitations of lower-dimensional models in capturing essential transport phenomena.
Notably, the simulations indicate a higher C\textsubscript{2} monomer selectivity of approximately 50~\% under plasma-based \itn-hexane pyrolysis, in contrast to the roughly 30~\% selectivity achieved via conventional fossil-fuel-based methods.
These findings underscore the potential advantages of plasma-driven pyrolysis and represent a critical step toward a comprehensive understanding of the complex thermochemical behavior governing plasma-assisted processes.

\end{minipage}
\end{center}


\doublespacing

\section{Introduction}

Climate change represents a significant challenge to contemporary society, necessitating immediate efforts to reduce carbon emissions in various sectors~\cite{ClimateChange0,ClimateChange1}. 
While various strategies have been proposed and implemented to reduce carbon emissions, electrification —a recent trend~\cite{electrification0} of replacing traditional fossil fuel-based energy sources with electricity—– stands out as one of the most promising approaches for achieving cleaner and more sustainable processes. Examples include electric vehicles and induction cooktops, which are already helping to reduce dependence on fossil fuels in transportation and household settings~\cite{Electrification4,Electrification5}.

\begin{figure}
\begin{center}
\includegraphics[width=0.6\textwidth]{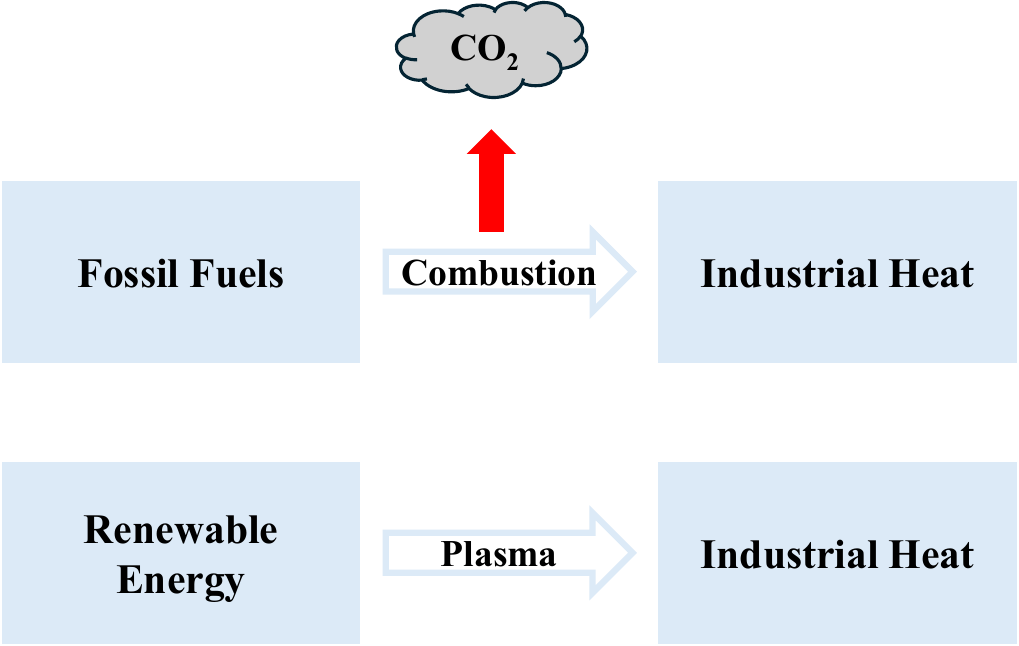}
\end{center}
\caption{Benefits of plasma as a clean heat source.
}\label{fig:plasmaheatsourcebenefit}
\end{figure}

Within the industrial sector, the petrochemical industry has significant potential to reduce carbon emissions by switching from fossil fuel-based energy sources to electricity, ideally derived from renewable sources~\cite{electrification1, electrification2, electrification3}.
This transition is especially promising in the context of the naphtha cracking process, a major contributor to carbon emissions within the sector. Naphtha cracking thermally decomposes complex hydrocarbons into key feedstocks such as ethylene, propylene, and other valuable olefins and aromatics~\cite{naphtha1}. 
This process comprises four key stages: pyrolysis, rapid quenching, compression, and separation and purification. These steps are heavily dependent on energy derived from the combustion of fossil fuels to achieve the high temperatures required for the cracking reactions. 
A promising strategy for the efficient electrification of this process is the use of plasma as a heat source during pyrolysis step~\cite{plasmaCracking}.
Unlike conventional heating, plasma-based pyrolysis achieves high temperatures by directly energizing the gas through electric fields, with plasma generated via arc discharge, RF, or microwave methods~\cite{JAHANMIRI2012416}. As shown in \fref{fig:plasmaheatsourcebenefit}, this approach has the potential to significantly reduce carbon emissions, because the electricity used to power the plasma can be sourced from various renewable energy options such as solar, wind, or hydroelectric power. 
There are additional advantages to using plasma as a heat source. Unlike conventional steam cracking, which transfers combustion heat through reactor walls, plasma generates heat directly within the reaction zone, thereby reducing thermal losses and improving overall energy efficiency.
Moreover, reactive species generated by the plasma may enhance reaction selectivity and increase the yield of desired products, making it a highly efficient and sustainable option for electrified pyrolysis in industrial applications.


These benefits collectively provide strong motivation for adopting plasma-based technologies for
electrified thermal processing. Various plasma systems such as arc discharge~\cite{glidingArc1, glidingArc2}, radio frequency (RF)~\cite{RF1, RF2, RF3}, and microwave-based reactors~\cite{microwave1, microwave2} have been explored for their ability to generate high-temperature environments suitable for hydrocarbon cracking~\cite{chanmi, cm1}. Although differing in power delivery mechanisms, these systems typically share a common structure: a plasma generation region, where the discharge gas (e.g., $\rm{H}_2$, Ar) is ionized by an electric field, and a downstream pyrolysis region, where the high-temperature plasma interacts with the hydrocarbon feedstock to initiate thermal decomposition. Especially, hydrogen is often employed in the discharge gas to suppress carbon formation and to promote a more uniform temperature distribution within the reactor, owing to its high thermal diffusivity~\cite{Hydrogen1}.

To investigate the fundamental pyrolysis behavior of naphtha cracking under these plasma conditions, \textit{n}-hexane is commonly used as a surrogate for naphtha because of its simpler molecular structure and well-characterized reaction kinetics. However, existing chemical reaction mechanisms for \textit{n}-hexane pyrolysis are largely developed for oxidative environments and involve an excessive number of species and reactions, making them unsuitable for simplified pyrolysis modeling under hydrogen-rich, non-oxidative conditions. 
This motivates us to develop a simplified yet accurate chemical reaction model to capture the key decomposition pathways of \textit{n}-hexane pyrolysis in a hydrogen plasma environment. 

\begin{figure}
\begin{center}
\includegraphics[width=0.7\textwidth, trim={3cm 3cm 3cm 3cm}, clip]{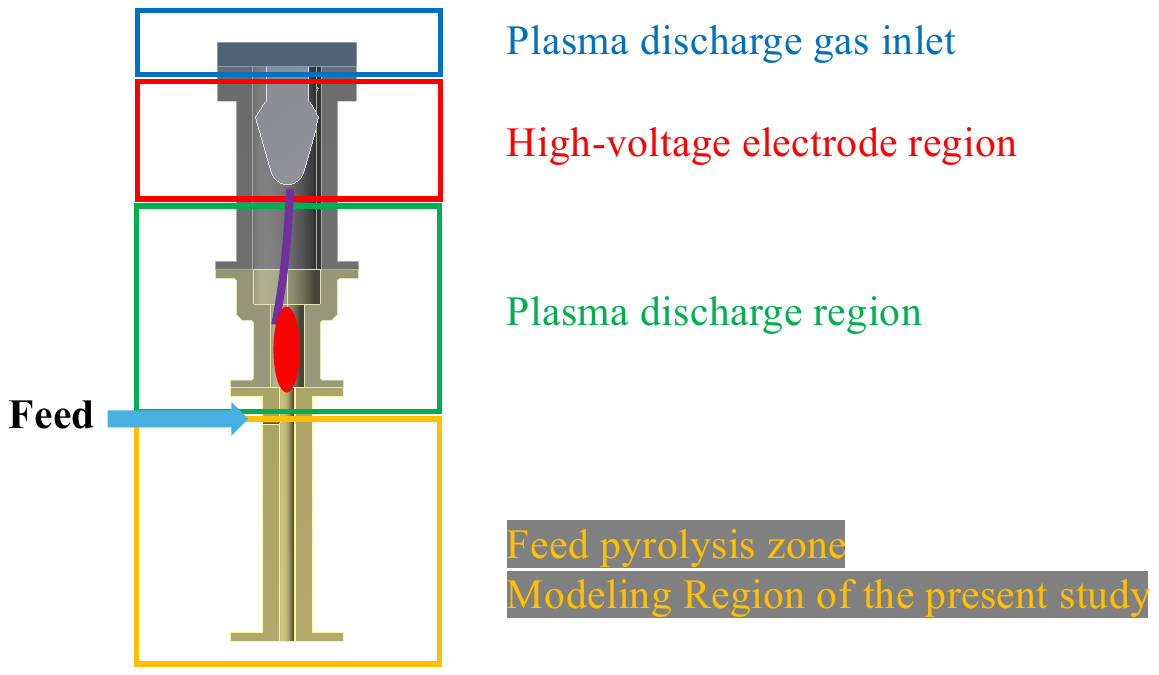}
\end{center}
\caption{Schematic of a arc plasma reactor.}\label{fig:reactorSchematic}
\end{figure}

A schematic of the reactor in our previous work \cite{chanmi, glidingArc1, glidingArc2} is shown in \fref{fig:reactorSchematic}. The system comprises four major components: (1) the discharge gas inlet (H₂, Ar, or their mixture), which supplies the plasma-forming gas; (2) the high-voltage electrode region, where ionization is initiated near atmospheric pressure; and (3–4) the regions facilitating downstream reaction and product extraction. While significant progress has been made through prior studies, optimizing plasma reactor design remains a significant challenge, as key parameters—such as reaction temperature, residence time, reactor geometry, and plasma gas composition—strongly influence conversion efficiency and product selectivity. Moreover, comprehensive experimental evaluation of these factors is time-consuming and cost-prohibitive, emphasizing the need for effective modeling and simulation tools in reactor development.

In this context, the primary goal of this study is to model and simulate the physical and chemical processes in the plasma reactor, focusing on the feed pyrolysis zone. To reduce computational complexity, the plasma is treated as a high-temperature gas mixture, which avoids the need for detailed plasma chemistry. Although the model is simplified, it still requires careful consideration of key process parameters such as reaction temperature, residence time determined by inlet gas velocity, reactor geometry, and plasma gas composition. Because experimental investigation of these factors is costly and time-consuming, a computational framework is developed to examine their effects on conversion and selectivity, and to support optimal reactor design. 

The rest of the paper is organized as follows. Section 2 introduces the full set of governing equations for modeling \textit{n}-hexane pyrolysis in the plasma reactor. 
To close these equations, a customized chemical kinetic mechanism tailored for the arc plasma reactor in \fref{fig:reactorSchematic} is developed using RMG~\cite{RMG1}, an open-source software that  automatically constructs detailed chemical kinetic mechanisms. 
In Section 3, we progressively build the analysis, starting with simplified 0-D and 1-D models to identify key design parameters 
such as feed gas flow rate and temperature. 
During this process, the chemical kinetic model is tested and validated.
Subsequently, computational fluid dynamics (CFD) analysis is performed using the developed mechanism, accounting for the effects of gas mixing and heat transfer to the environment.
This is followed by a detailed discussion of the simulation results and potential strategies for improving pyrolysis efficiency.
Section 4 concludes the paper with a summary and recommendations for future work.

\section{Method}

We begin by outlining our overall approach to model the pyrolysis of \textit{n}-hexane under high temperature condition provided in the arc plasma reactor. The quantity of interest (QoI) is chosen as the conversion rate and selectivity of the $n$-hexane pyrolysis.
To effectively calculate the Quantity of Interest (QoI), 
we focus on the chemical and physical mechanisms occurring within the feed pyrolysis zone of the reactor.
The two main tools are
the chemical reaction mechanism model and  CFD with the energy balance equation (i.e., heat transfer).

\subsection{Formulation}
\label{subsec:formulation}

\begin{figure}
\begin{center}
\includegraphics[width=0.45\textwidth]{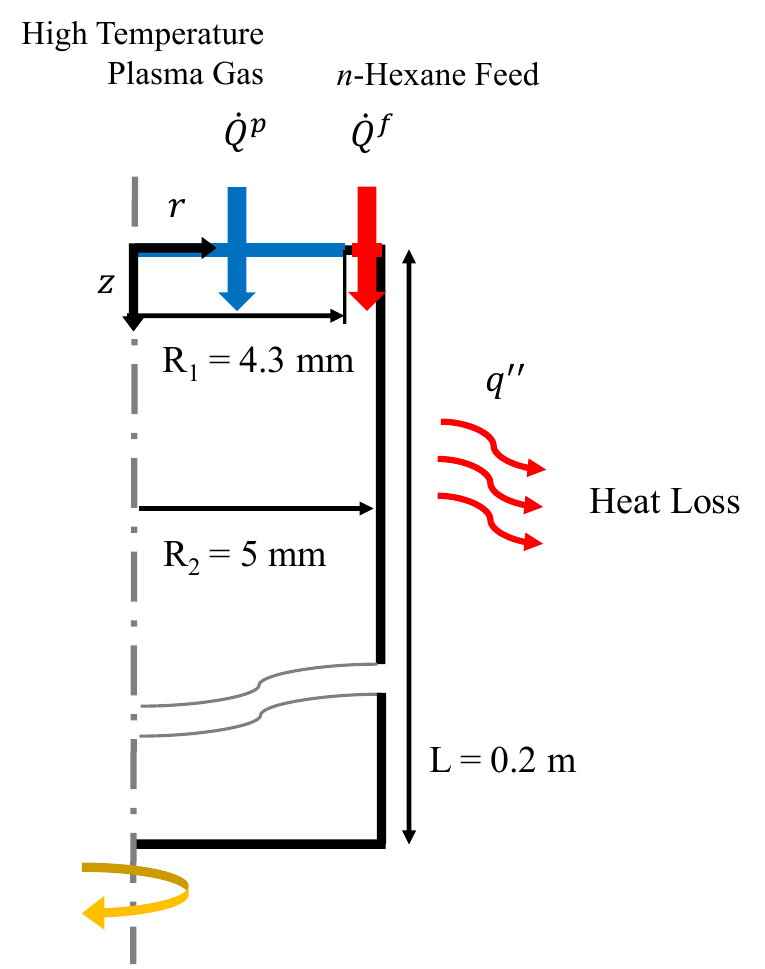}
\end{center}
\caption{The axisymmetric problem domain of this study focuses on the interior regions of the plasma reactor, which are enclosed by stainless steel (SUS304). The grey dash-dot line denotes the axis of symmetry.
}
\label{fig:problemdomain}
\end{figure}

The feed pyrolysis zone of the plasma reactor, visualized in \fref{fig:problemdomain}, includes the interior regions of the plasma reactor, which are enclosed by stainless steel SUS304.
The surrounding environment is at atmospheric pressure and room temperature ($300\; \rm{K}$).

The feed zone has an inner radius $R$ of $0.005\; \rm{m}$ and a length of $L=0.2 $ m. The volumetric flow rate (of H\textsubscript{2} and Ar plasma discharge gas mixture) $\dot{Q}^p$ is $35\; \rm L/min$ under normal conditions ($\rm{NLPM}$; $298.15\; \rm{K}$, $1\; \rm{atm}$). 
Considering the expansion of the gas mixture at high temperatures approaching $2,273\; \rm{K}$,
the average flow velocity is estimated to be approximately $\bar u^p = 76.60 \; \rm m/s$. 
On the other hand, \itn-hexane feed gas is injected through the wall with a flow rate 
$\dot{Q}^f$ estimated as only a order of $2\; \rm m/s$ ($0.01\; \rm{NLPM}$). 
The Reynolds number ($\mathrm{Re}=\rho \bar{u^p} D/ \mu$) of the gas flow is approximately estimated by 3,000. 
This estimation is based on the ideal gas law and Sutherland's law~\cite{Sutherland} to approximate the density $\rho$ and dynamic viscosity $\mu$ of the gas mixture at high temperatures, $T>1,000\;\rm K$.

We employ the formulations of chemically reacting fluid flow with total $N$-participating chemical species. The plasma discharge gases,  are assumed to be incompressible and to behave as ideal gases. 
The set of governing equations is written as follows:

\begin{itemize}
    \item Mass conservation for mixture gasses:
    \begin{equation}
        \frac{\partial \rho }{\partial t} + \nabla \cdot(\rho \bu) =0
    \label{eqn:total_mass_conservation}
    \end{equation}
    \item Mass conservation for chemical species (also known as species transport equations):
    \begin{equation}
        \frac{\partial \rho Y_i}{\partial t} + \nabla \cdot(\rho  Y_i \bu) = - \nabla \cdot \mathbf{j}_i + \mathsf{R}_i \quad (i=1,2,...N)
    \label{eqn:species_transport}
    \end{equation}
    \item Momentum conservation  
    \begin{equation}
    \frac{\partial \rho \bu }{\partial t} + \nabla \cdot ({\rho \bu \otimes \bu)}= -\nabla p + \nabla \cdot \bm{\tau} + \rho \sum^{N}_i Y_i \mathbf{f}_i
    \end{equation}
    \item The conservation of the total energy 
    \begin{equation}
       \rho\mathsf{C}_p \left( \frac{\partial T}{\partial t} + \bu \cdot \nabla T \right)  = \left(
        \frac{\partial p}{\partial t}+
        \bu \cdot \nabla p 
       \right)
       -\nabla \cdot \mathbf{q}
      - \bm{\tau} : \nabla \bu 
      - \sum^{N}_i h_i \mathsf{R}_i.
      \label{eqn:energy_equation}
    \end{equation}
    where the enthalpy $h$ and static specific heat at constant pressure $\mathsf{C}_p$ of gas mixture are formed in terms of those for each species by 
    \begin{equation*}
        h=\sum^{N}_{i=1}Y_i h_i.
    \end{equation*}
    and
    \begin{equation*}
        \mathsf{C}_p=\sum^{N}_{i=1}Y_i \mathsf{C}_{p\; i}.
    \end{equation*}
    
    \item Thermodynamic Equation of State:
    \begin{equation}
        p = \rho \mathcal{R} T M^{-1}, \quad M =\left( \sum^{N}_{i=1} Y_i/M_i \right)^{-1}.
        \label{eqn:equation_of_state}
    \end{equation}
\end{itemize}

In the above set of equations, $\rho$ means the density, $\bu$ denotes the velocity vector, and $p$ stands for the pressure. Also, $h$ represents enthalpy, $T$ the temperature, and $\mathcal{R}$ the universal gas constant. 
$Y_i$ stands for the mass fraction,
$\mathbf{f}_i$ denotes the body force per unit mass of chemical species $i$.  
$M$ stands for the mean molecular mass, while $M
_i$ means the molecular mass of species $i$.
Also, $\bm{\tau}$ is the viscous tensor, 
$\mathbf{j}_i$ the diffusive flux vector of species, 
$\mathbf{q}$ the thermal flux vector respectively, of which forms are
summarized in Appendix~\ref{app:variousForms}.
Lastly,  
$\mathsf{R}_{i}$, the net rate of production of species $i$ by chemical reaction, will be mainly discussed in \sref{subsec:rmg}.
To ensure consistency with mass conservation, the species mass fractions, diffusion velocities, and chemical source terms must adhere to the following conditions:
\begin{equation}
    \sum^{N}_i Y_i=1, \quad 0\leq Y_i \leq 1
\end{equation}
\begin{equation}
    \sum^{N}_i \mathbf{j}_i=0
\end{equation}
\begin{equation}
    \sum^{N}_i \mathsf{R}_i=0.
    \label{eqn:sum_of_R}
\end{equation}

The mass conservation equations~\eref{eqn:total_mass_conservation}-\eref{eqn:species_transport} require the mass fractions $Y_i$ and total mass flow rate $\dot m$ at the inlet of a computational domain. 
These are typically are typically calculated from the total volumetric flow rate $\dot{Q}$ and the molecular fractions $C_i$ of the each species,
and applying the ideal gas law~\eref{eqn:equation_of_state} to calculate $\rho$. 
\begin{equation}
    \dot m = \rho \dot{Q}
\end{equation}
\begin{equation}
    Y_i =\frac{C_i M_i}{\sum_j C_j M_j}. 
\end{equation}
Also, note that summation of conservation equations for all species in Eq.~\eref{eqn:species_transport} implies total mass conservation~\eref{eqn:total_mass_conservation}, 
so that one of those N + 1 equations is redundant.
Additionally, solving the energy equation~\eref{eqn:energy_equation} requires appropriate thermal boundary conditions, such as a heat transfer coefficient $h$, to account for heat transfer to the environment.

The QoI of this study, in determining $\dot{Q}^f$, is  
\begin{equation}
    \mathrm{Conversion \; of\;} n-\mathrm{hexane} =  \frac{\dot{n}^{inlet}_{\rm{n-}\rm{hexane}}-\dot{n}^{outlet}_{\rm{\itn-hexane}}}{\dot{n}^{inlet}_{\rm{\itn-hexane}}}
\end{equation}
and 
\begin{equation}
    \mathrm{Selectivity \; of \; i-species} = \frac{\mathcal{C}_{\rm{i}}\dot{n}^{outlet}_i}{\mathcal{C}_{\rm{i}}\dot{n}^{converted}_{\rm{\itn-hexane}}}
    \label{eqn:selectivity}
\end{equation}
where $\mathcal{C}_{\rm{i}}$ is the number of carbon atom and $\dot{n_i}$ is molar flow rate of $i$-species, which is calculated as
\begin{equation}
    \dot{n_i} = \frac{\dot{Q}C_i}{\dot{Q}^m},
\end{equation}
where $\dot{Q}^m$ is the molar volume flow rate.

The following section is devoted to describe a chemical mechanism model
to provide information on the rate of production for each species $\mathsf{R}_i$ in \eref{eqn:species_transport}, 
which is one of important milestones of the present study. 
It requires a detailed, step-by-step description of the underlying chemical transformations that occur during pyrolysis. This will close the formulation summarized in this section. 

\subsection{Development of Chemical Kinetics Mechanism using RMG}
\label{subsec:rmg}

From the viewpoint of chemical kinetics and statistical mechanics, the pyrolysis of \itn-hexane proceeds through a step-by-step sequence of elementary reactions, which refer to the indivisible steps of a chemical reaction.\footnote{In general, elementary reactions are equipped with well-defined rate laws, with the reaction rate being proportional to the reactant concentrations raised to powers equal to their stoichiometric coefficients~\cite{Espenson}.
} 
Such a set of elementary reactions that collectively describe the step-by-step process of a chemical transformation is called a chemical mechanism. 
It comes with information on the formation and consumption of intermediate species.
Once such mechanism for \itn-hexane pyrolysis is determined, 
an explicit form of the net mass production rate $\dot{ \mathsf{R}}_i$ is provided (see Appendix~\ref{app:productionRate}), and the formulations in the previous section become complete.  

There exist several reaction mechanisms for describing the pyrolysis of \itn-hexane in literature~\cite{nhexaneMechansm_1,nhexaneMechansm_2,nhexaneMechansm_3}. 
Unfortunately, however,  
such existing \itn-hexane pyrolysis mechanism cannot be directly employed to the present study because of the following reasons:
\begin{enumerate}
    \item These reaction mechanisms were originally developed for oxygen-rich environments, such as combustion process, whereas the plasma reactor for hydropyrolysis process operates under a hydrogen-based atmosphere with negligible or no oxygen present.
    \item They assume reaction mechanisms operate under relatively low-temperature conditions ($T< 1,000\; \rm{K}$), resulting in low sensitivity to high-temperature environment of plasma reactors, which often exceed $1,000\; \rm{K}$. 
    \item Existing mechanisms are large-scale kinetic models with thousands of reactions and hundreds of species, primarily designed for oxygen-rich environments. However, since oxygen is absent under our pyrolysis conditions, many of these reactions are unnecessary, leading to unnecessarily high computational costs in CFD analysis.
\end{enumerate}

Therefore, a new reduced-order chemical kinetic mechanism is needed to represent reactions under high-temperature, hydrogen-rich, and low-oxygen conditions. 
In the past, kinetic models are manually constructed by tracking species and reactions, which is time-consuming, error-prone, and requires expert knowledge.
To circumvent such issues, we employ an open-source software package called, the Reaction Mechanism Generator (RMG)~\cite{RMG1, RMG2, RMG3, RMG4, RMG5}, to automate the construction of  
a chemical kinetic model for pyrolysis of \itn-hexane under the unique environment of the plasma reactor. 

A set of user inputs of RMG is reaction pressure, temperature and molecular fractions of initial species $C_i$.
Then, it relies on database of reaction rate coefficients
and executes core species enlarging algorithm to determine important elementary reactions using the specific flux tolerance which is also given by a user.
Note that the size (the number of reactions and chemical species) of the chemical kinetic model is implicitly determined by flux tolerance given to the algorithm.
We refer readers to the recent review~\cite{RMG1} for more detailed background on the underlying theories and computational algorithms for these automated procedures. 

From Eq.~\eref{eqn:species_transport}, it is clear that the number of partial differential equations that need to be solved during the CFD analysis strictly increases with N, the number of chemical species in chemical kinetic model. 
For practical reasons, a compromise must be made between model simplicity and the high accuracy potentially achievable with a large number of chemical reactions and species. In this context, the following section begins with the validity of the kinetic models employed in this study.

\section{Results}

In this section, we systematically explore predictions using the RMG chemical kinetic model, beginning with the simplest configuration. Each analysis is conducted with a specific objective:

\begin{itemize}
    \item 0-D Analysis: Validation of the RMG model; identifying the conversion rate and product selectivity trends in terms of reaction temperatures under varying residence times.
    \item 1-D Analysis: Identifying product selectivity at different positions within the reactor.
    \item CFD Analysis: Considering reactor geometry, heat transfer, and gas mixing effects.
\end{itemize}

\subsection{0-dimensional analysis}

In 0-D analysis, we consider the simplest model where the spatial dimensions of the system are neglected; 
the system is treated as a single point or as having no spatial variation. 
Also, chemical reactions are assumed to occur in a well-mixed environment where all properties, such as concentration and temperature, are uniform throughout the system. The concentration of each species is only a function of time. 
The formulations prescribed in \sref{subsec:formulation} reduces to a set of ordinary differential equations
\begin{equation}
    \frac{dX_i}{dt}= \mathsf{R}_i
    \label{eqn:0d_chem_ode}
\end{equation}
where $X_i(= MM^{-1}_i Y_i)\;  \rm [mol/kg]$ is the mole concentration of species. 

The first goal of 0-D-analysis is to validate the RMG-generated chemical mechanism model using the existing data: LLNL(n-Heptane)~\cite{LLNL}: Lawrence Livermore National Laboratory, JetSurF(Version 2.0)~\cite{JetSurF} and the experimental results of Ref.~\cite{exp}.

\begin{table}
\begin{center}
\begin{tabular}{ccc} 
\hline
Quantity & RMG-test  & RMG \\
\hline 
Pressure   & 1 atm & 1 atm   \\
Temperature     & 1,273 K & 1,373 K \\
$C_{\rm{Ar}}$ & 0.996 & 0.463 \\
$C_{\rm{H_2}}$ & 0 & 0.463 \\
$C_{\rm{n-hex.}}$ & 0.004 & 0.073  \\
Maximum Carbon number & 6  & 6\\ 
Termination Conversion & 0.9 & 0.9\\
Termination Time & $10^6$ s & $10^6$ s\\ 
\hline
\end{tabular}
\caption{The initial conditions (i.e., the input parameters) for the RMG were used to generate the set of reaction mechanisms for mechanism validation. 
}
\label{tab:0d_RMGinput1}
\end{center}
\end{table}

To this end, the chemical mechanisms are generated based on the RMG initial conditions summarized in Table~\ref{tab:0d_RMGinput1}.
With this RMG input script, total 125 reactions of 32 dominant species are created. This model is hereafter referred to as RMG-test model. The number is significantly reduced compared to that of LLNL and Jetsurf; LLNL has total 2,450 reactions of 550, and JetSurf has 2,163 reactions of 348 species. 
This result is unsurprising, given that most of the reduction arises from the exclusion of reactions and species involving oxygenated compounds.

\begin{figure}
   \centering
   \subfloat[Species profiles of \itn-hexane and $\rm{C}_2\rm{H}_4$]{ 
    \includegraphics[width=0.7\linewidth,trim={0cm 0cm 0cm 0cm},clip]{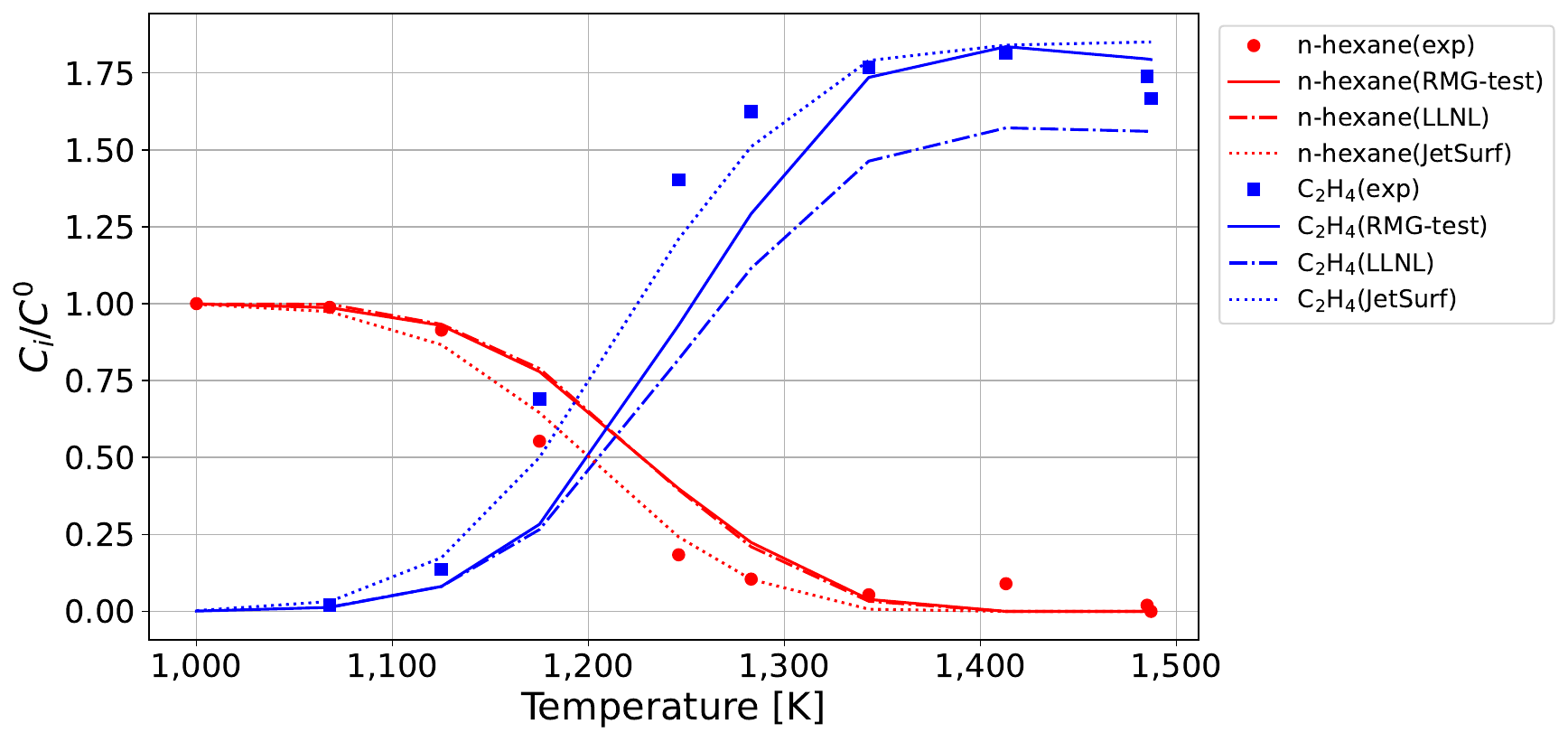}}\\
   \subfloat[Species profiles of $\rm{C}_2\rm{H}_2$ and $\rm{C}_2\rm{H}_6$]{
   \includegraphics[width=0.7\linewidth,clip]{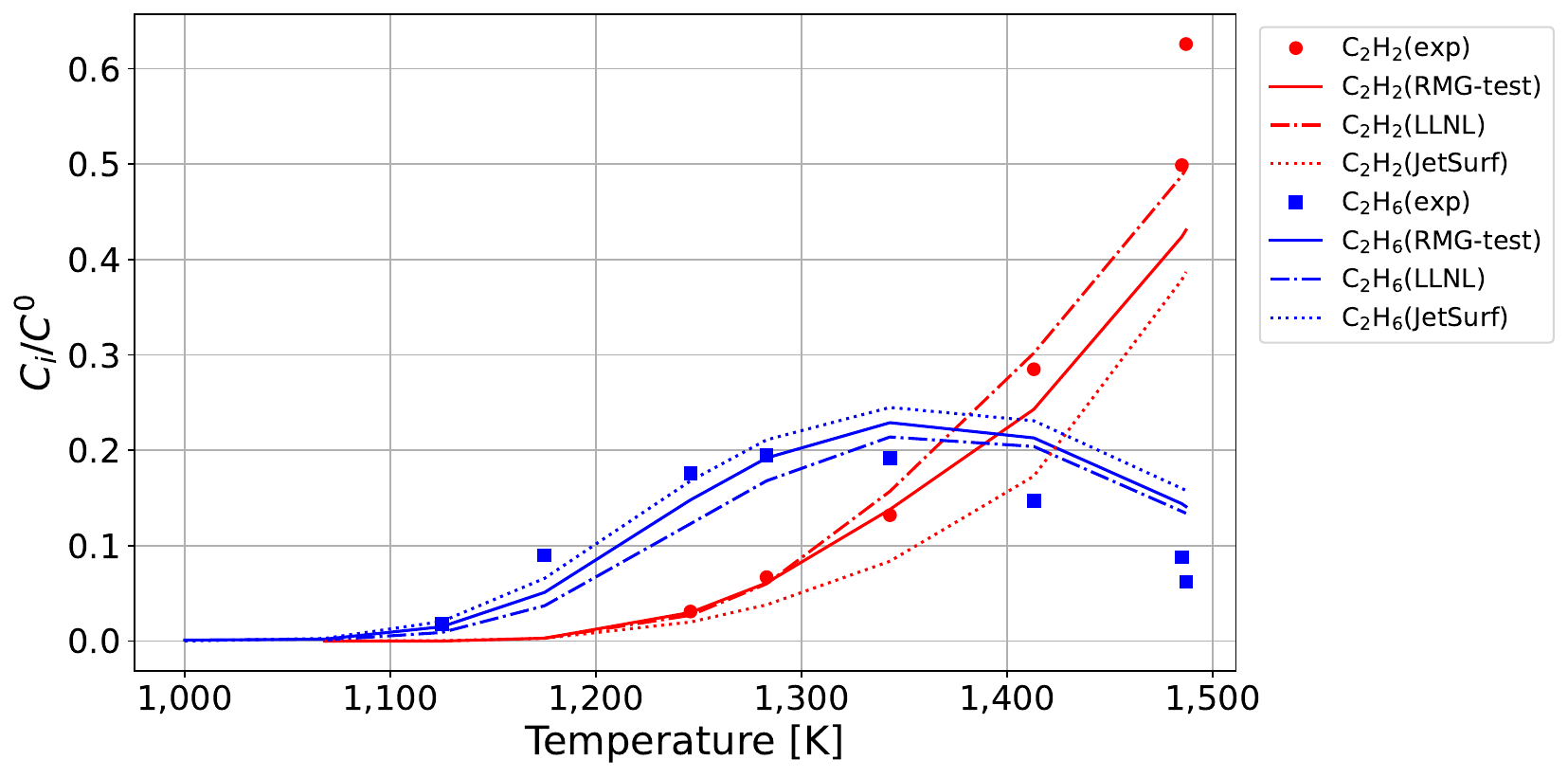}}
   \caption{Comparison of normalized species concentration profiles between the single-pulse shock tube experiment and 0-D simulation results using different chemical models: RMG-test, LLNL, and JetSurF.}
   \label{fig:0d_result_comparison}
\end{figure}

To validate RMG-test model and other chemical mechanisms for analyzing pyrolysis of \itn-hexane, 
we first collected experimental measurements of the yields of various products during \itn-hexane pyrolysis from the literature. 
Specifically, in Ref~\cite{exp}, 
the mole fraction of primary species (\itn-hexane, C\textsubscript{2}H\textsubscript{4},
C\textsubscript{2}H\textsubscript{2}, and
C\textsubscript{2}H\textsubscript{6})
at various reaction temperatures ranging from $1,000\; \rm{K}$ to $1,500\; \rm{K}$ are reported through the shock-pulse tube test, a conventional experimental technique used to investigate high-temperature chemical kinetics under rapid heating. 
The reaction time $t_R$ in these tests varies with the reaction temperature. For temperatures of $1,000,\; 1,100,\; 1,200,\; 1,300,\; 1,400$ and $1,500\; \rm{K}$, $t_R$ was set to 1,820, 1,750, 1,670, 1,600, 1,500 and 1,500 microseconds, respectively.
Then, the mole fractions $C_i$ of the primary species, normalized by the initial concentration $C^0$ of \itn-hexane is marked with with red circles and blue squares in plots of \fref{fig:0d_result_comparison}.

Next, the closed homogeneous batch reactor model in \texttt{Chemkin-Pro}~\cite{CHEMKIN} is used to solve the set of ODEs~\eref{eqn:0d_chem_ode}
with LLNL, JetSurf and RMG-test mechanism. 
The temperatures and reaction times are set to match the experimental conditions, while the initial time step size of the solver was set to $\Delta t= 10^{-5}\; \rm{s}$, which was then adaptively adjusted during the simulation.
The results are presented using line plots: dash-dot for LLNL, dashed for JetSurf, and solid for the RMG-test, respectively in \fref{fig:0d_result_comparison}.
Overall, the RMG-test model, despite employing a significantly reduced number of chemical reactions and species, effectively captures the yield trends across a range of temperatures. Thus, the RMG-test model demonstrates reasonable accuracy, making it a practical and efficient choice for simulating \itn-hexane pyrolysis in the subsequent CFD analysis.

\begin{figure}
   \centering
   \subfloat[ $t_R=1 \;\rm ms$ ]{ 
    \includegraphics[width=0.5\linewidth,trim={0cm 0cm 0cm 0cm},clip]{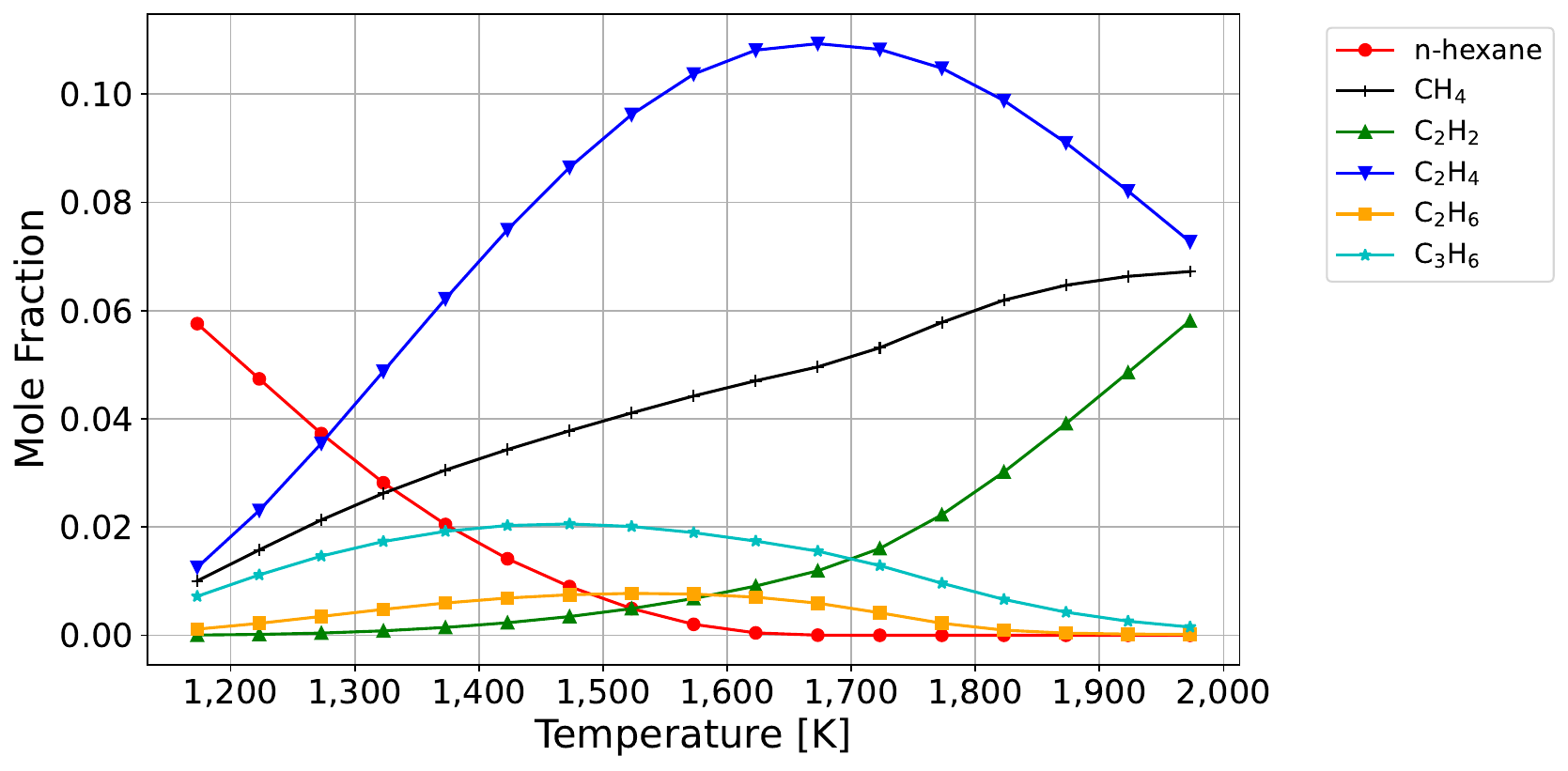}
    \label{subfig:0d_result_comparison1_a}
    }
   \subfloat[ $t_R=10 \;\rm ms$ ]{
   \includegraphics[width=0.5\linewidth,clip]{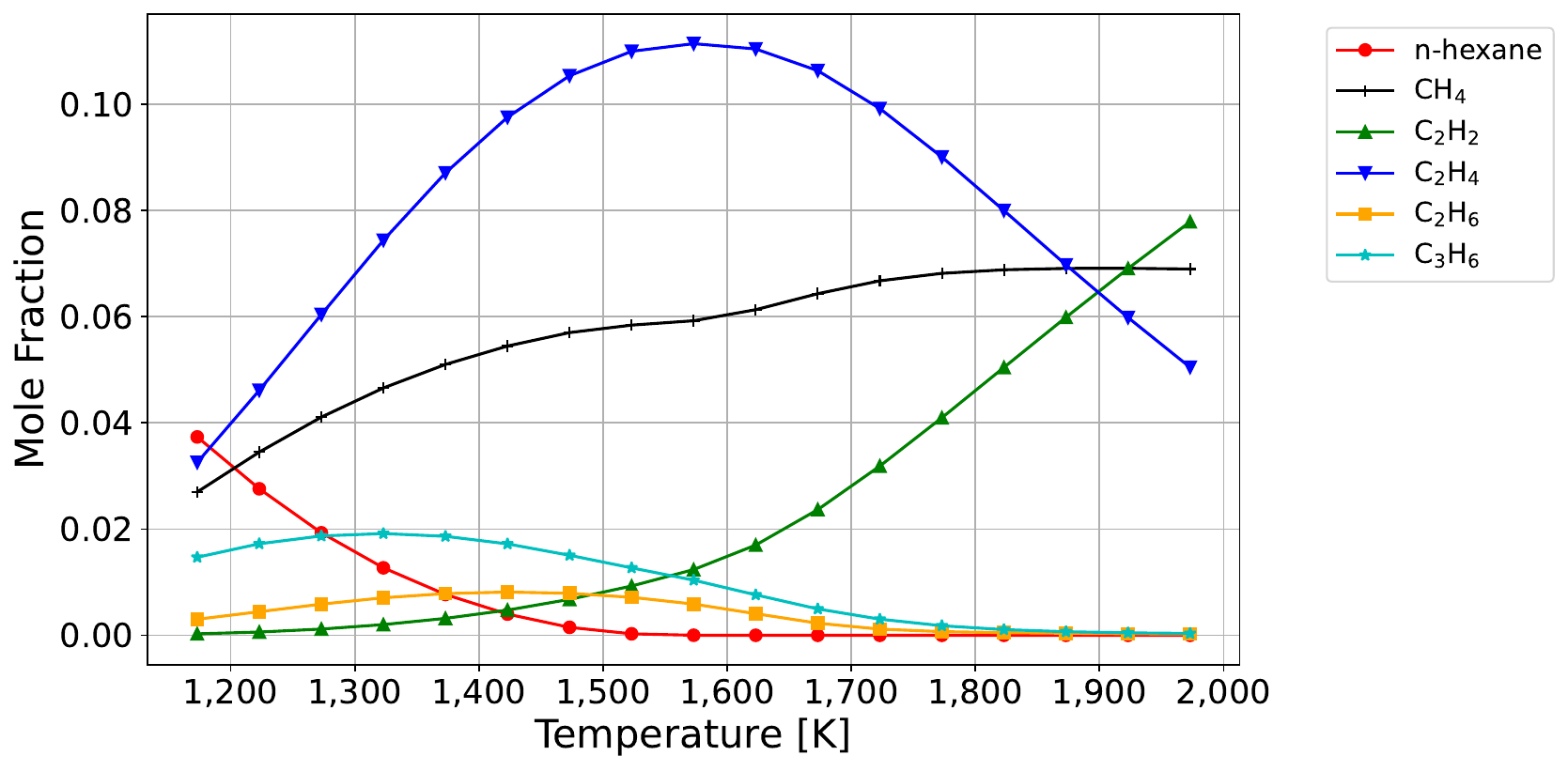}
   \label{subfig:0d_result_comparison1_b}
   }
   \\
   \vspace{0.5cm}
    \subfloat[ $t_R=100 \;\rm ms$ ]{
    \includegraphics[width=0.5\linewidth,clip]{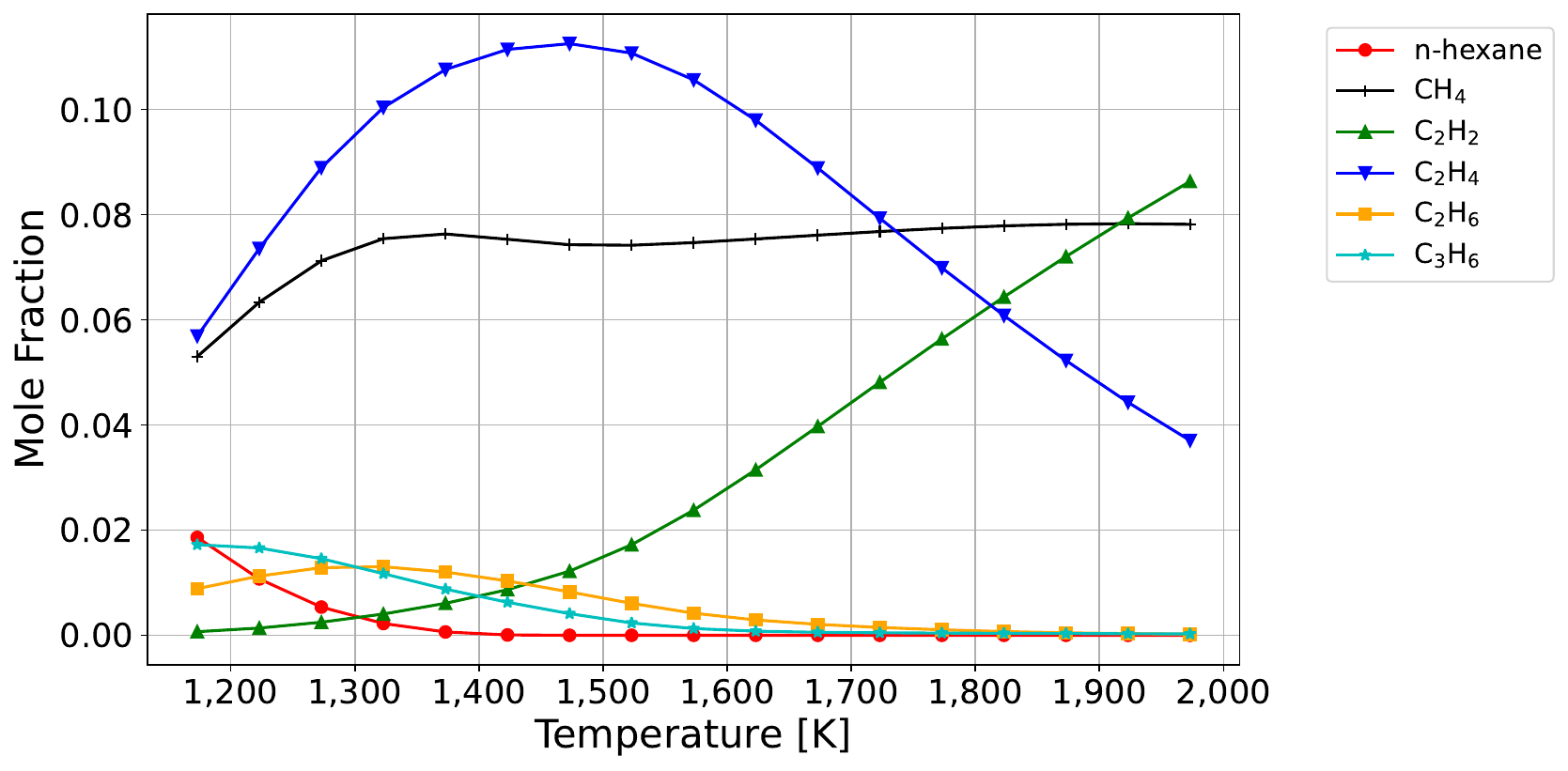}
    \label{subfig:0d_result_comparison1_c}
    }
   \subfloat[ $t_R\to \infty$]{
   \includegraphics[width=0.5\linewidth,clip]{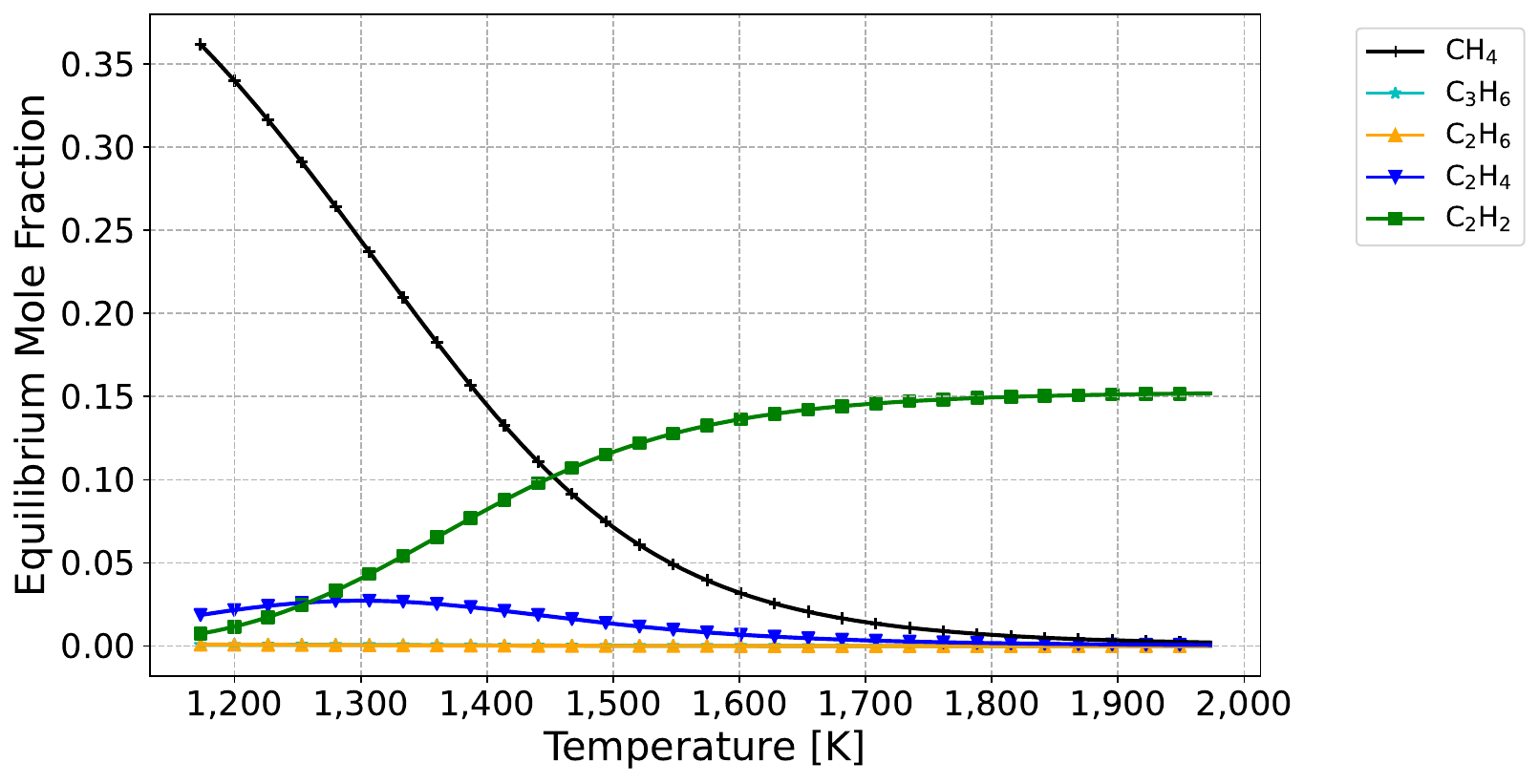}
   \label{subfig:0d_result_comparison1_d}
   }
   \\
   \caption{
   0-D analysis of \itn-hexane pyrolysys. Temperature-dependent mole fractions of products at various reaction times $t_R$.
   }
   \label{fig:0d_result_comparison1}
\end{figure}

After validating the RMG-test chemical mechanism model, we now introduce a new mechanism specifically developed for pyrolysis reactions under plasma conditions.
This model is hereafter referred to as the RMG model.
The initial conditions that were used to generate the model is again summarized in Table~\ref{tab:0d_RMGinput1}, which include equal initial concentrations of Ar and H\textsubscript{2}. 
Next, using the RMG model, we re-investigate the yields of various products in different temperature ranges $1,173-1,973\; \rm{K}$ at atmospheric pressure, with the results summarized in \fref{fig:0d_result_comparison1}.

These results suggest optimal reaction temperatures and times for potentially maximizing the selectivity of the target product C\textsubscript{2}H\textsubscript{4}, 
represented by the blue line in \fref{fig:0d_result_comparison1}a-d. 
For instance, at longer reaction time ($t_R\to\infty)$, the maximum mole fraction of C\textsubscript{2}H\textsubscript{4} remains around 0.03, 
which is significantly lower than at shorter reaction time, where it exceeds 0.1. Meanwhile, the mole fraction of other products, such as C\textsubscript{2}H\textsubscript{2}, increase with the reaction times.
This is because C\textsubscript{2}H\textsubscript{4} acts as an intermediate during the pyrolysis of \itn-hexane. At higher temperatures and long reaction times, further decomposition or dehydrogenation of C\textsubscript{2}H\textsubscript{4} into smaller species such C\textsubscript{2}H\textsubscript{2}. 
To achieve high selectivity of C\textsubscript{2}H\textsubscript{4},
a temperature range $1,300-1,500\; \rm{K}$ appears necessary, 
particularly for short reaction times ($t_R< 100 \; \rm{\mu s}$).
A shorter reaction time is also generally preferable, provided that conversion and selectivity are maintained, as it contributes to improved reactor efficiency.
From a practical standpoint, it is important to maintain the reaction temperature well below $1,700\; \rm{K}$ to ensure the safe and sustainable operation of stainless-steel (SUS) reactors, whose melting point is approximately $1,700\; \rm{K}$. 

That said, these results alone are not sufficient to determine the optimal operating conditions. Real plasma reactors involve additional complexities, such as spatial variations and the mixing dynamics between plasma discharge gases and \itn-hexane —factors that are beyond the scope of the present 0-D analysis. Therefore, these findings will be used as auxiliary references to inform subsequent reactor design, and we proceed to a 1-D analysis to address these additional considerations.

\subsection{One-dimensional analysis}

\begin{figure}
   \centering
   \subfloat[Distance dependent mole fractions of products at a reactant flow rate of $10\; \rm{mL/min}$]{ 
    \includegraphics[width=0.8\linewidth,trim={0cm 0cm 0cm 0cm},clip]{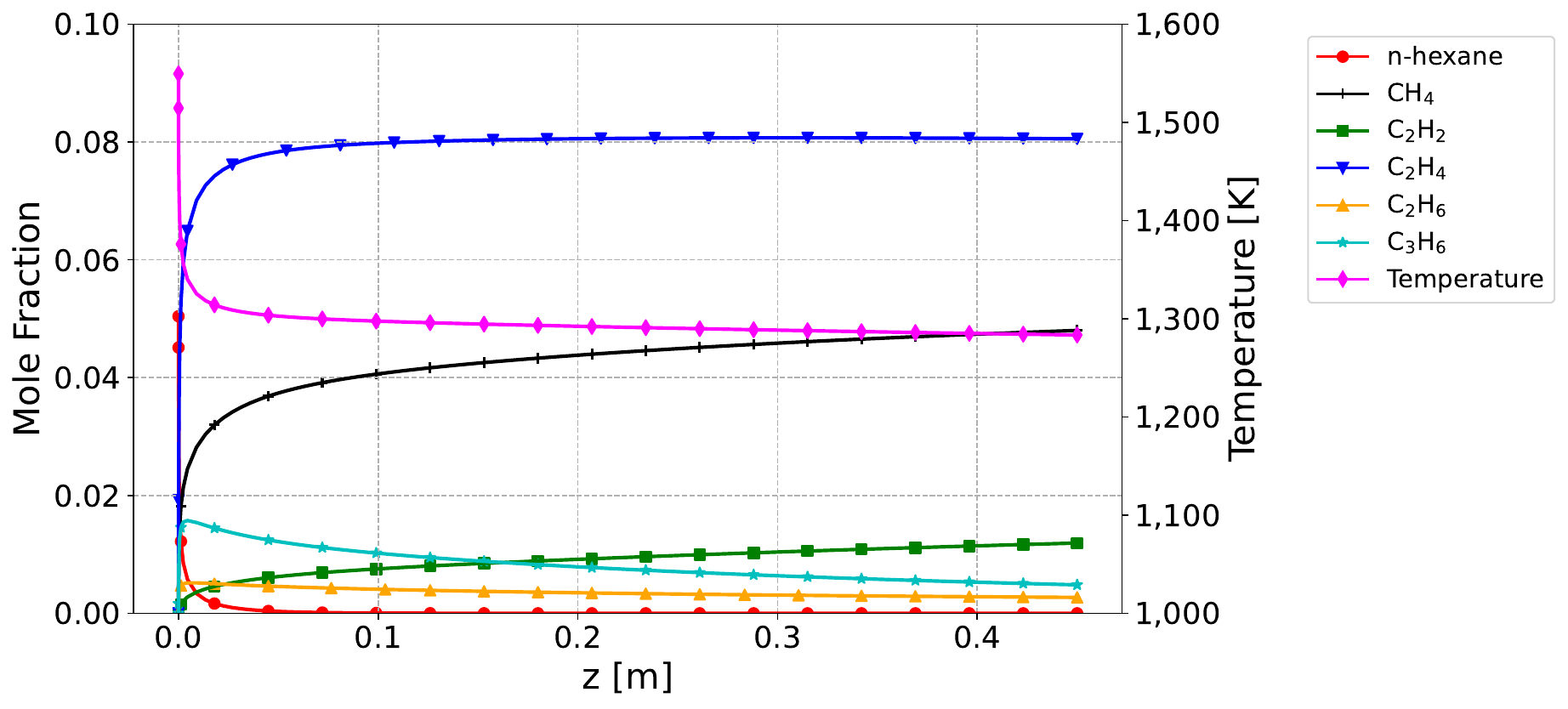}}
    \\
    \subfloat[Distance dependent mole fractions of products at a reactant flow rate of $15\; \rm{mL/min}$]{ 
    \includegraphics[width=0.8\linewidth,trim={0cm 0cm 0cm 0cm},clip]{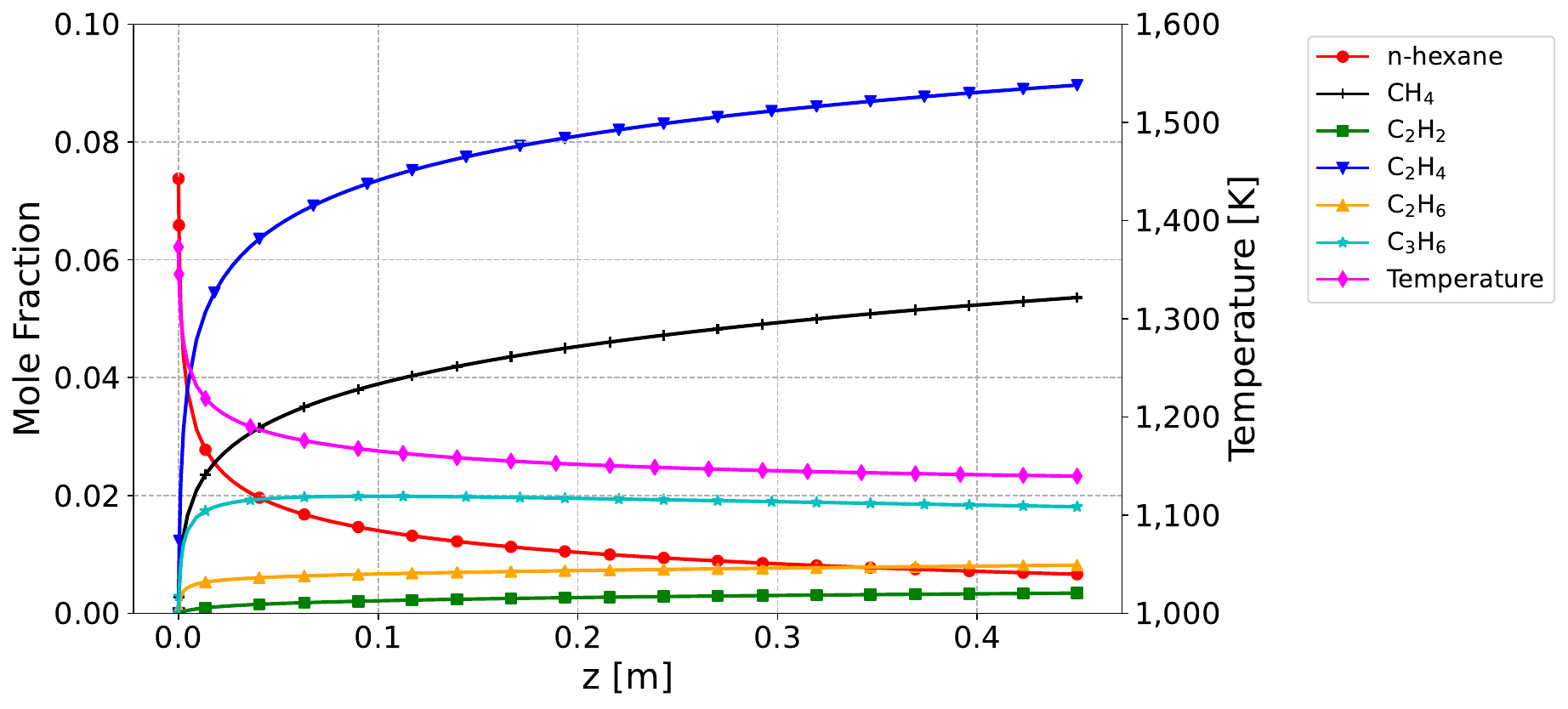}}
    \\\subfloat[Distance dependent mole fractions of products at a reactant flow rate of $20\; \rm{mL/min}$]{ 
\includegraphics[width=0.8\linewidth,trim={0cm 0cm 0cm 0cm},clip]{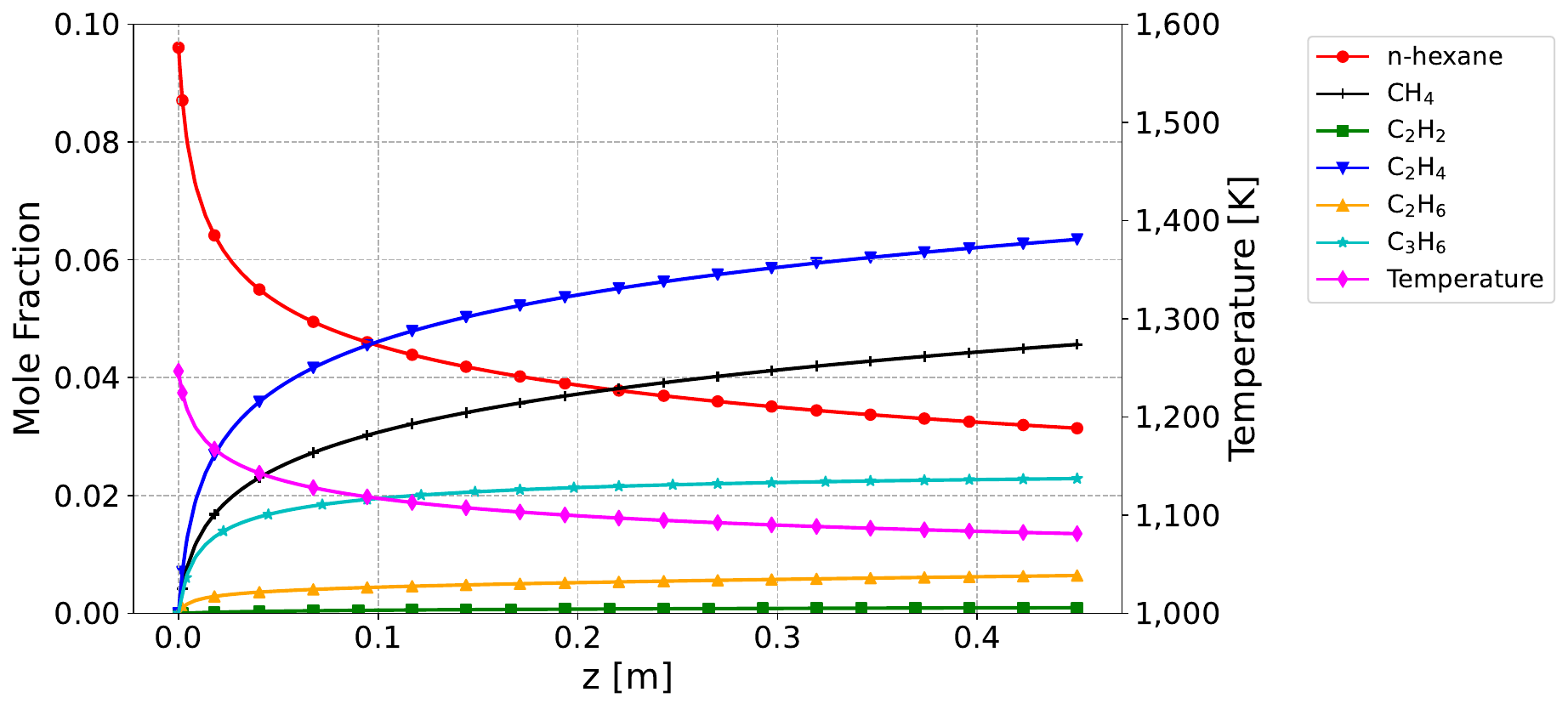}}
    \\
   \caption{
   1-D analysis of \itn-hexane pyrolysis at various \itn-hexane flow rates. Plasma discharge gas (Ar: $17.5\; \rm{NLPM}$, H$_2$:$17.5\; \rm{NLPM}$) is supplied at $2,273\; \rm{K}$, and \itn-hexane at $300\; \rm{K}$.
   }
   \label{fig:1dAnalysis} 
\end{figure}

\begin{figure}
   \centering
   \subfloat[Distance dependent product selectivity and reactant conversion at reactant flow rate of $10\; \rm{mL/min}$]{ 
    \includegraphics[width=0.8\linewidth,trim={0cm 0cm 0cm 0cm},clip]{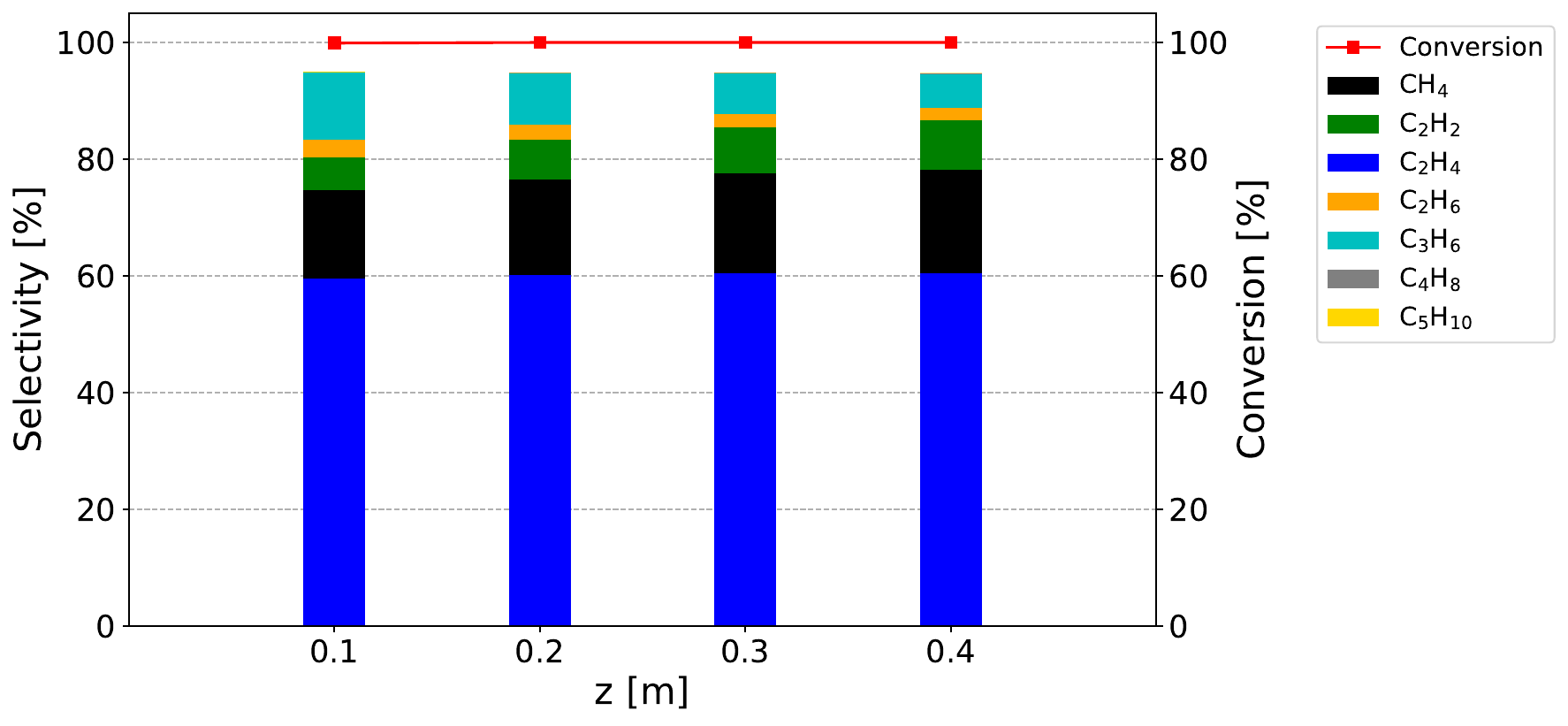}}
    \\
    \subfloat[Distance dependent product selectivity and reactant conversion at reactant flow rate of $15\; \rm{mL/min}$]{ 
\includegraphics[width=0.8\linewidth,trim={0cm 0cm 0cm 0cm},clip]{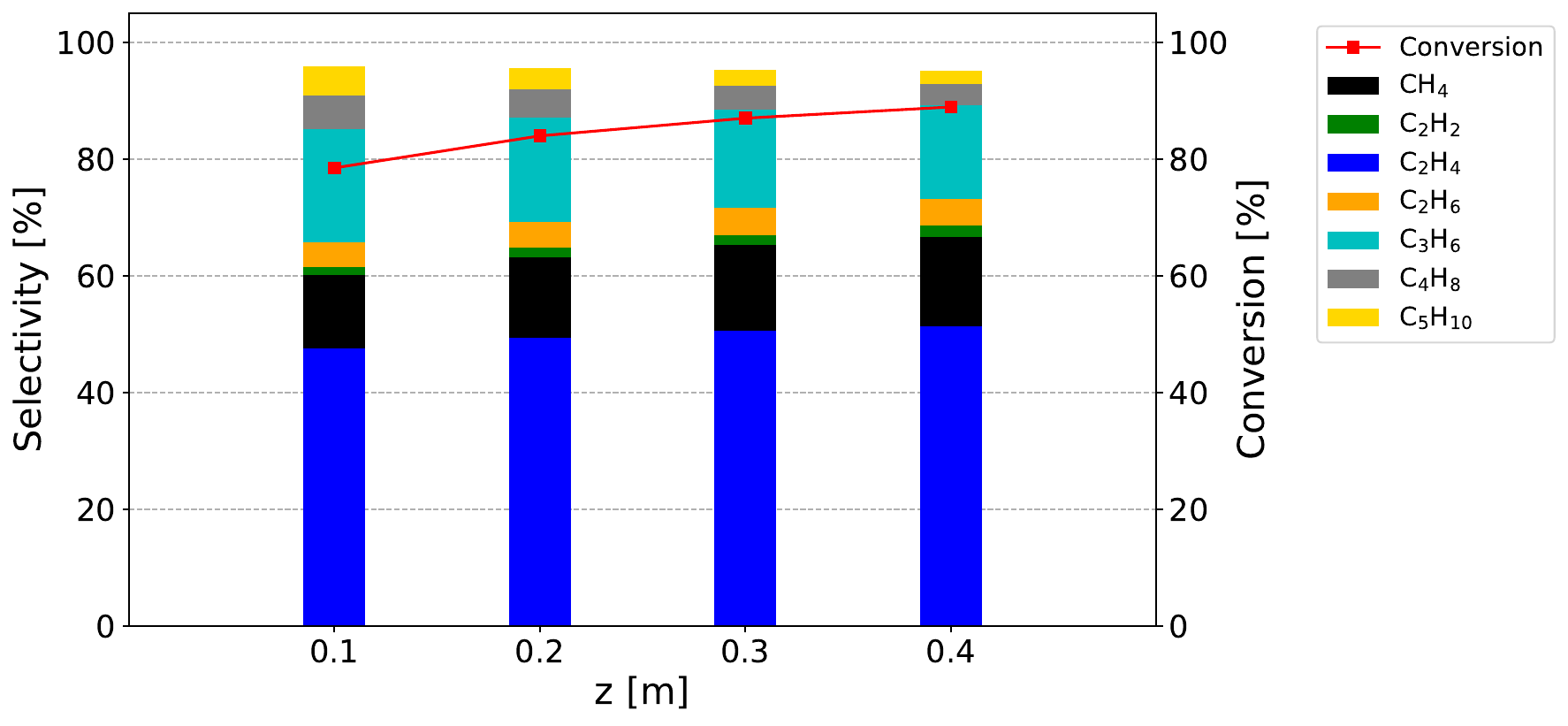}}
    \\
    \subfloat[Distance dependent product selectivity and reactant conversion at reactant flow rate of $20\; \rm{mL/min}$]{ 
    \includegraphics[width=0.8\linewidth,trim={0cm 0cm 0cm 0cm},clip]{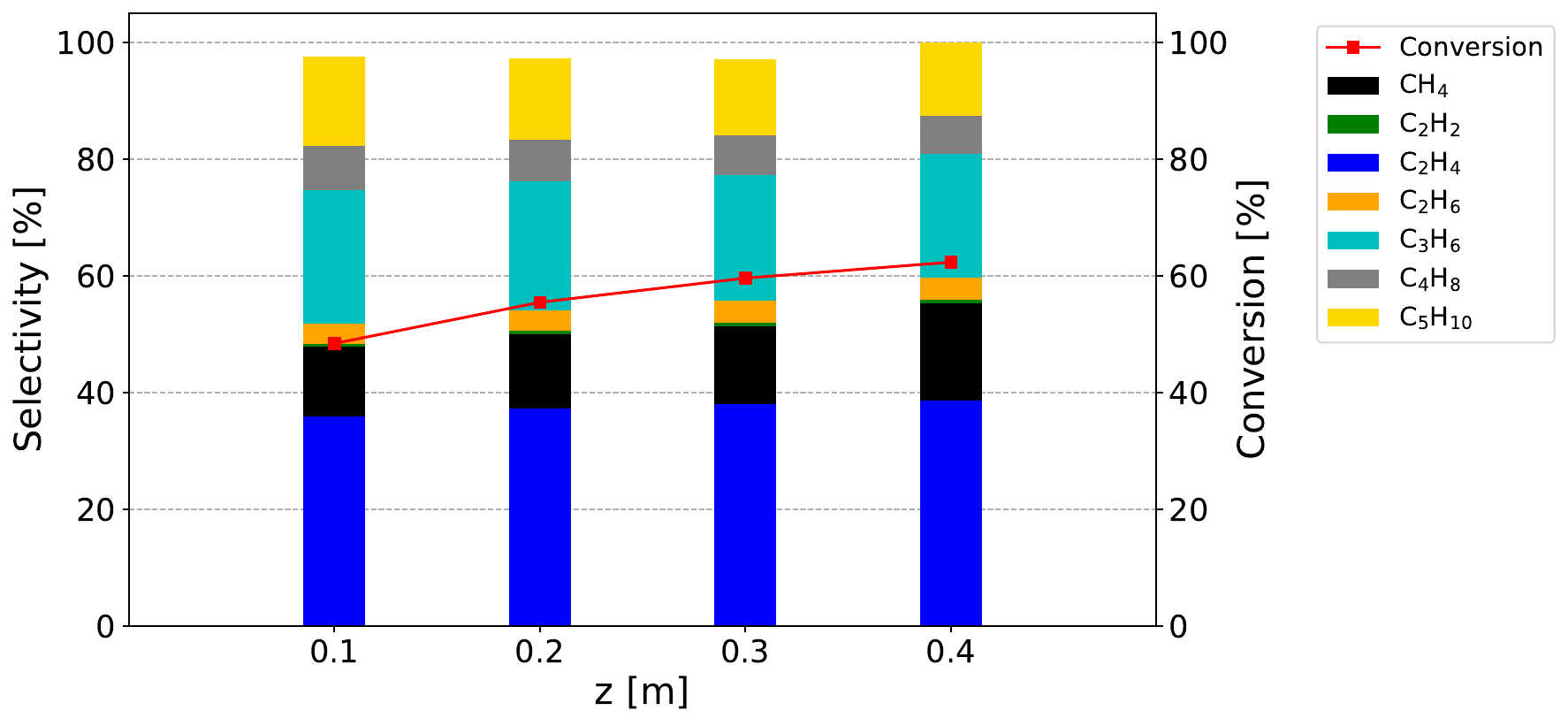}}
    \\
    \caption{
   Selectivity and conversion along the reactor length from 1-D analysis at different reactant flow rate.
   }
   \label{fig:1Danalysis_Graph} 
\end{figure}

The primary objective of 1-D analysis is to understand how chemical species and flow properties evolve spatially along the reactor axis under realistic operating conditions.
Although reaction temperature and residence time, which are mainly studied in 0-D analysis, are essential factors for optimizing chemical conversion, they are not directly controlled when operating the pyrolysis reactor. 
Instead, given the plasma feed gas composition and reactor geometry (i.e., diameter $D=0.01\; \rm{m}$), these parameters become functions of the \itn-hexane feed flow rate $\dot{Q}^f$. Therefore, understanding how temperature and species concentrations vary along the main flow direction in the reactor is critical for informing reactor design and performance prediction.

To this end, CHEMKIN~\cite{CHEMKIN} 1-D analysis is performed using the Plug Flow Reactor (PFR) solver to predict the production of primary product species as a function of the \itn-hexane feed gas rate. 
The assumptions underlying the PFR solver are summarized as follows:
\begin{itemize}
    \item Steady, unidirectional flow 
    \item No axial diffusion (no back-mixing).
    \item Perfect radial mixing (uniform properties across any cross-section).
\end{itemize}

These assumptions allow the velocity of the plasma and feed gas mixture to be determined solely from the continuity equation~\eref{eqn:total_mass_conservation}. On the other hand, during 1-D analysis, 
the energy equation~\eref{eqn:energy_equation} is solved under adiabatic boundary conditions, with no heat exchange considered between the system and its environment.
The plasma discharge gas, consisting of Ar ($17.5\; \rm{NLPM}$) and H\textsubscript{2} ($17.5\; \rm{NLPM}$), is assumed to enter the reactor at a temperature of $2,273\; \rm{K}$, fully mixed with the feed gas at $300\; \rm{K}$.
This mixing is achieved using an ``ideal mixer element'' in CHEMKIN, which performs complete mixing without chemical reactions.  
As no heat loss or reaction occurs during the mixing process, the temperature of the resulting homogeneous gas mixture is determined solely by the enthalpy balance between the two gas streams.
This mixture serves as the inlet condition for the subsequent 1-D pyrolysis simulation.

The results of the 1-D analysis for three different feed gas flow rates ($\dot{Q}^{f}= 10,\; 15,\; 20 \; \rm mL/min$) are shown in \fref{fig:1dAnalysis} and \fref{fig:1Danalysis_Graph}. First, \fref{fig:1dAnalysis} shows the spatial profiles of temperature and the mole fractions of the main products along the reactor length.
In all cases, a steep temperature drop is observed along the flow direction, driven by the progression of the endothermic reaction.
The mole fractions of all products increase along the length of the reactor.
However, since mole fraction does not directly represent selectivity—the primary quantity of interest in this analysis, as defined in \eref{eqn:selectivity}—we further summarize the selectivities of major products and conversion of \itn-hexane at four key reactor positions
($z=0.1,\; 0.2,\; 0.3$ and $0.4\; \rm{m}$) in \fref{fig:1Danalysis_Graph}.

At $\dot{Q}^f=10\; \rm mL/min$, complete conversion of \itn-hexane is achieved throughout the reactor, indicating capacity for higher feed rates without sacrificing conversion.
The mole fraction of C\textsubscript{2}H\textsubscript{4} stabilizes at 0.08 beyond $z = 0.1\; \rm{m}$, with selectivity reaching approximately 60~\%. Additionally, a notable trend is observed where C\textsubscript{2}H\textsubscript{2} selectivity increases with distance, eventually exceeding that of C\textsubscript{3}H\textsubscript{6} at later positions. This shift implies enhanced secondary cracking reactions of heavier species into lighter species as the residence time increases along the flow direction. 

In the case of $\dot{Q}^f=15\; \rm mL/min$, the conversion is not complete, but progressively increases from approximately 80~\% at $z=0.1\; \rm m$ to approximately 90~\% at $z=0.4\; \rm m$. While C\textsubscript{2}H\textsubscript{4} is still the dominant product, its selectivity is reduced compared to the $\dot{Q}^f=10\; \rm mL/min$ case. This can be explained by the lower average temperature along the reaction zone due to higher feed rates, which slows down the formation of lighter products such as C\textsubscript{2}H\textsubscript{4} and C\textsubscript{2}H\textsubscript{2}. Instead, relatively higher selectivity is observed for heavier hydrocarbons such as C\textsubscript{3}H\textsubscript{6}, C\textsubscript{4}H\textsubscript{8}, and C\textsubscript{5}H\textsubscript{10}. As the flow progresses, the selectivity of these heavier species decrease, while that C\textsubscript{2}H\textsubscript{4} increases, reflecting gradual progress of thermal cracking reactions. 

An increased feed flow rate, $\dot{Q}^f=20\; \rm mL/min$, results in a temperature of the pyrolysis zone below $1,100\; \rm{K}$. Consequently, this lower temperature environment coupled with a short residence time not only leads to conversion remaining below 60~\% even at $z = 0.4\; \rm{m}$, but also to a significant suppression in the formation of C\textsubscript{2}H\textsubscript{4}. These results suggest that a feed rate above $\dot{Q}^f=15\; \rm mL/min$ exceeds the capacity of the system to sustain efficient pyrolysis.

In summary, these findings highlight the sensitivity of the product distribution to thermal conditions and residence time
—both of which are directly influenced by the feed flow rate— emphasizing the need for its precise control in pyrolysis system optimization. 
Although the simplicity of the one-dimensional model limits these results and prevents full representation of a real plasma reactor. they still offer useful guidance for narrowing design parameters, such as flow rate, before running more complex CFD simulations.

\subsection{CFD analysis}

The goal of CFD analysis is to understand how (a) heat transfer with the environment and (b) mixing between the plasma discharge gas and the \itn-hexane feed gas affect conversion and selectivity in the pyrolysis reactor.
Unlike the idealized 1-D model—which assumes complete mixing without reactions—actual reactor conditions involve finite mixing between two gas streams with significantly different temperatures ($2,273\; \rm{K}$ and $300\; \rm{K}$). This leads to non-equilibrium thermal states that affect the reaction process.
To explore how these effects influence pyrolysis characteristics, this section is divided into two parts. The first compares 1-D and CFD results, highlighting the additional impact of mixing and heat transfer captured only in CFD. 
The second isolates the individual effects of mixing and heat transfer within the CFD framework to quantify their separate contributions.

\subsubsection{Comparison Between 1-D Analysis and CFD Simulation}

For the CFD analysis, the governing equations~\eref{eqn:total_mass_conservation}-\eref{eqn:sum_of_R} were numerically solved within a two-dimensional axisymmetric domain shown in \fref{fig:problemdomain}. 
The set of equations is solved using the pressure-based solver in the commercial software \texttt{fluent} 2024 R1~\cite{Fluent}. 
In particular, the following settings were applied. First, the chemical reactions involved in the plasma-assisted pyrolysis occur on extremely short time scales, typically in the order of microseconds.
Such reactions exhibit strong stiffness due to the presence of both fast and slow chemical time scales, which can lead to numerical instability.
Thus, to solve the species transport equations~\eref{eqn:species_transport},  we adopted a stiff chemistry solver in Fluent, which employs implicit integration schemes that ensure stable and accurate solutions even under rapid reaction kinetics. 
The ``Volumetric reactions'' option captures reaction rates and heat/mass generation within the reactor volume, while the ``Diffusion Energy Sources'' option accounts for energy from reactant/product diffusion, ensuring accurate temperature and heat transport profiles. These settings are essential for simulating fast reaction dynamics and energy distribution.

\begin{figure}
\begin{center}
\includegraphics[width=0.8\textwidth]{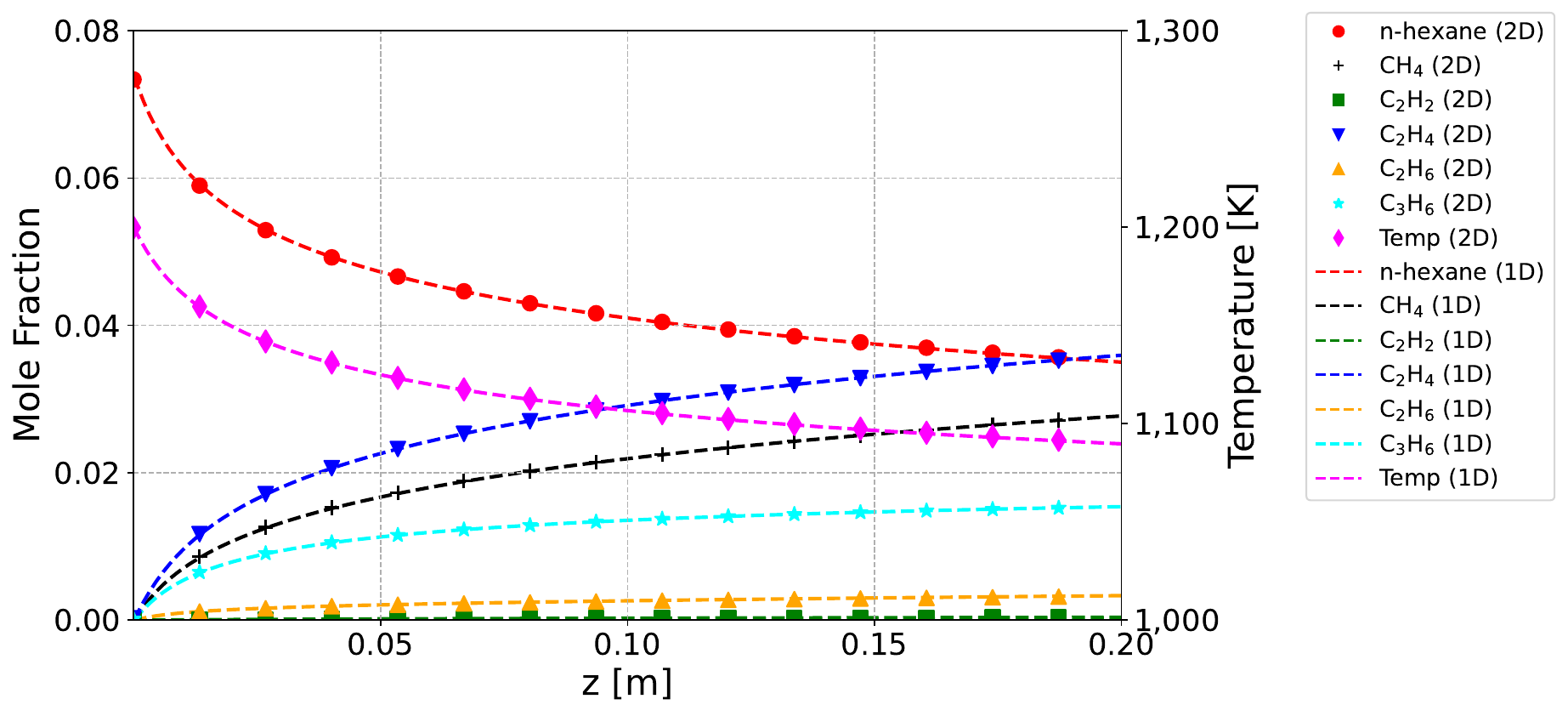}
\end{center}
\caption{Comparison of temperature and mole fractions of primary species between the quasi-1-D CFD simulation (2-D domain with 1-D settings) and the PFR simulation in Chemkin-Pro.
}\label{fig:cfd_validation}
\end{figure}

To validate the prescribed CFD settings, 
we compare the temperature and mole fraction of primary species of the quasi-1-D CFD simulation in a rectangular domain (solving a 2-D domain but with actual 1-D settings) with those from the 1-D PFR reactor solver in Chemkin-Pro. The result is shown in \fref{fig:cfd_validation}. In the plot, the former is represented by dots, while the latter is denoted by dashed lines.
The close agreement between the two confirms the adequacy of the present CFD setup.

On the other hand, the main CFD analysis considers a cylindrical reactor with an axisymmetric geometry. 
A cylindrical coordinate system $(r,z)$ is used, where $z$ denotes the axial direction. 
The origin is defined at the point where the centerline intersects the entrance region (see \fref{fig:problemdomain}). 
To model heat loss, a wall heat transfer coefficient of $20\; \rm{W/(m^2\cdot{K})}$ was applied, with the value referenced from Ref.~\cite{heattransfer1} as a representative estimate for general cases.
Next, a non-premixed configuration was simulated by spatially separating the inlets of the plasma discharge gas and the reactant: the plasma discharge gas, a mixture of Ar and $\rm{H}_2$ ($17.5\; \rm{NLPM}$ each, therefore $\dot{Q}^{p}= 35\;  \rm NLPM$) at $2,273\; \rm{K}$, was introduced through the central inlet of the reactor. The reactant was injected near the wall at $300\; \rm{K}$ and the analysis was repeated for three different feed flow rates: $\dot{Q}^{f}= 10,\; 15$ and $20\; \rm mL/min$. 
The RNG $k-\epsilon$ model~\cite{kepsilon1} is used to account for the turbulent effects of the high-speed gas mixture flow. 
The flow domain is discretized by approximately 25,000 meshes, while the unknown variables $p,\bu, T, Y_i$ are discretized using the second order upwind scheme. 
The simulations were executed on 48 cores (2.9 GHz) with 528 GB of RAM.
It took approximately 4 hours to obtain a steady state solution. 

\begin{figure}
   \centering
   \subfloat[ 1-D ]{
   \includegraphics[width=0.7\linewidth,clip]{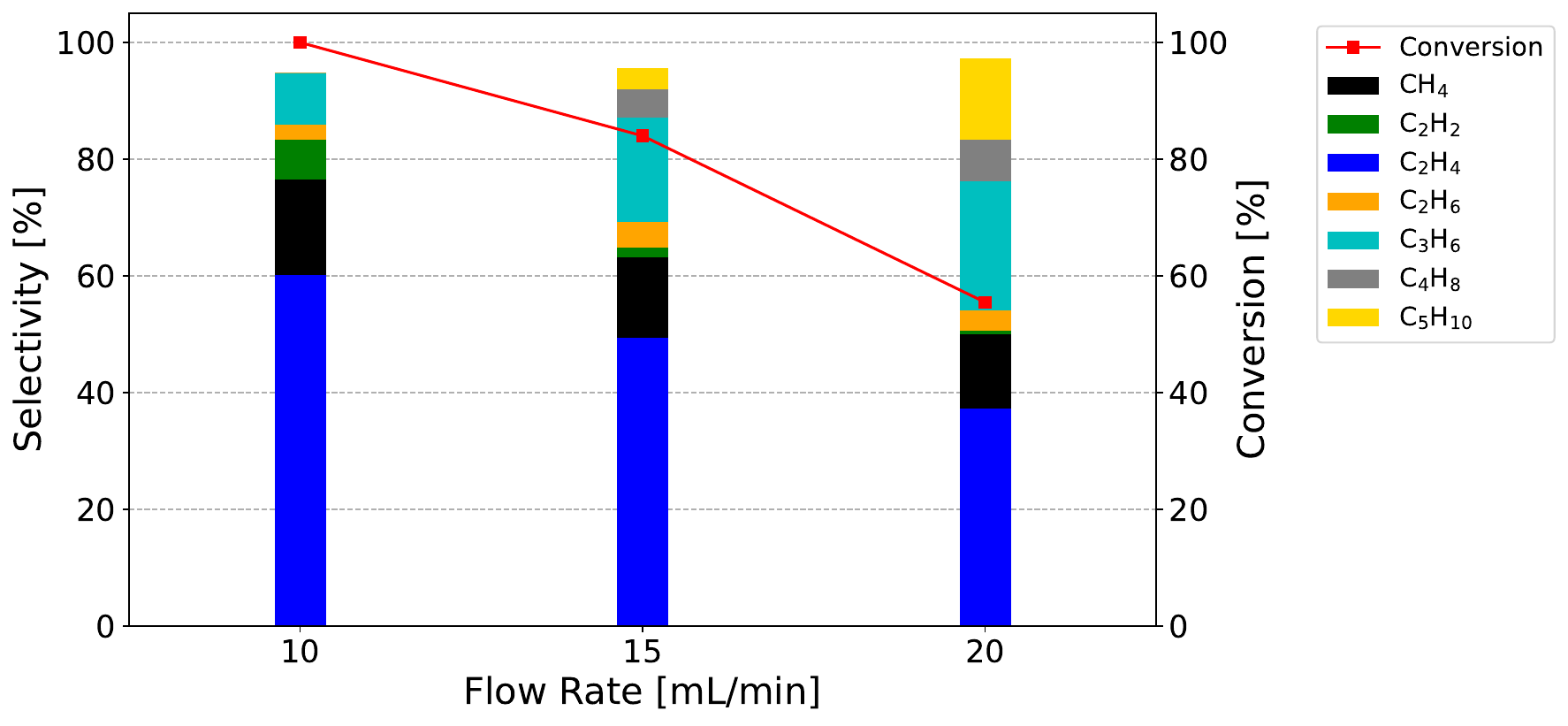}
   \label{subfig:CFDanalysis_Graph_a}}
   \\
   \subfloat[ CFD ]{ 
    \includegraphics[width=0.7\linewidth,trim={0cm 0cm 0cm 0cm},clip]{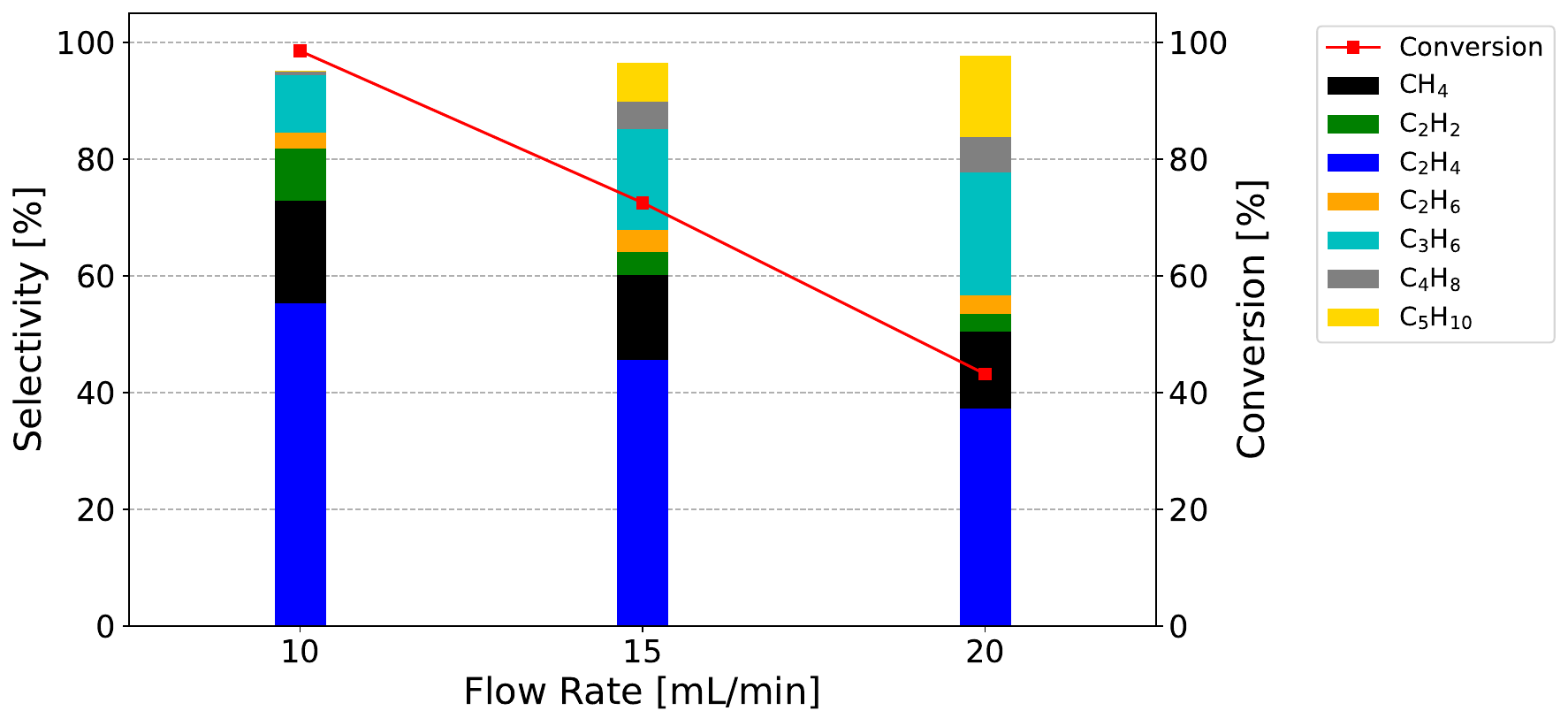}
    \label{subfig:CFDanalysis_Graph_b}
    }
    \\
   \caption{1-D and CFD results for product selectivity and \itn-hexane conversion at different feed flow rates.
   }
   \label{fig:CFDanalysis_Graph}
\end{figure}

The selectivity and conversion results from the previous 1-D CFD analysis for three different feed flow rates ($\dot{Q}^f= 10,\ 15,\ 20 \; \rm mL/min$) are compared in \fref{subfig:CFDanalysis_Graph_a} and \fref{subfig:CFDanalysis_Graph_b}. 
We observe that while the conversion remains unchanged at $\dot{Q}^f= 10 \; \rm mL/min$, 
it decreases by 10-20~\% at $\dot{Q}^f= 15,\ 20 \; \rm mL/min$.  
The selectivity of the target product, C\textsubscript{2}H\textsubscript{4},
also decreases by approximately 5~\% at $\dot{Q}^f= 10,\ 15\; \rm mL/min$, 
while it remains nearly unchanged at $\dot{Q}^f= 20\ \rm mL/min$. 
The results suggest that the 1-D analysis overall overestimates both the conversion of \itn-hexane and the selectivity of C\textsubscript{2}H\textsubscript{4}. 
From this, we also conclude that $\dot{Q}^f= 10\ \rm mL/min$ is the maximum feed capacity of the reactor that maintains 100~\% \itn-hexane conversion without compromising its performance.

While the CFD analysis corrected the overestimation of C\textsubscript{2}H\textsubscript{4} yield observed in the 1-D analysis, the predicted value still exceeds 50~\%. Notably, this remains significantly higher than the typical maximum yield of 31~\% achieved in conventional naphtha cracking processes~\cite{ethylene1}.
The results highlight the superior ethylene productivity of the plasma-assisted pyrolysis system.
In contrast, the C\textsubscript{3}H\textsubscript{6} yield remains comparable to that of the conventional process, 
at around 15~\%, 
further demonstrating that the enhanced ethylene yield does not come at the cost of propylene selectivity.
To better understand the origins of this enhanced performance, we next examine the individual roles of mixing and heat transfer.

\begin{figure}
\begin{center}
\includegraphics[width=0.9\textwidth]{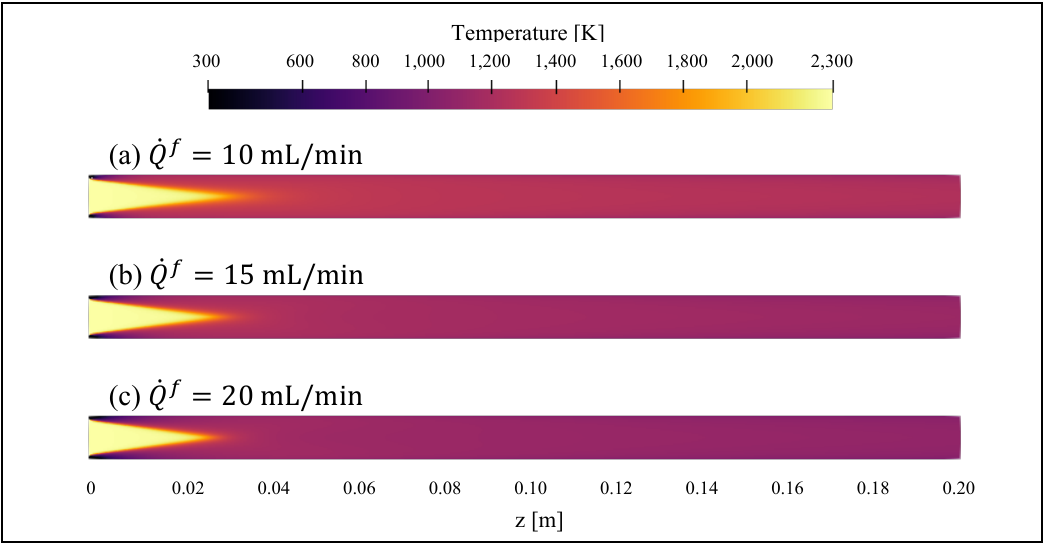}
\end{center}
\caption{Temperature contour profiles under varying \itn-hexane flow rates. 
At the inlet $(z=0)$, the feed gas and plasma discharge gas remain unmixed, and the heat loss coefficient at the reactor wall is set to 20~[$\rm W/(m^2\cdot K)$].
}
\label{fig:CFD_T_flow}
\end{figure}

\begin{table}
\begin{center}
\begin{tabular}{c|c|c|c|c} 
\hline
Condition & Case A  & Case B & Case C & Full CFD \\
\hline
Premixed & O & O & X & X \\
Heat transfer coefficient at wall $h$ [$\rm W/(m^2\cdot K)$] & 0 & 20  & 0 & 20 \\
\hline 
Conversion [\%]   & 99.57   &  98.04   &  99.93   &  98.53  \\
Sel. of CH\textsubscript{4} [\%] & 17.23 & 16.52 & 18.30 & 17.57 \\
Sel. of C\textsubscript{2}H\textsubscript{2} [\%] & 9.15 & 7.24 & 10.15 & 8.88 \\
Sel. of C\textsubscript{2}H\textsubscript{4} [\%] & 60.64 & 60.47 & 55.47 & 55.32 \\
Sel. of C\textsubscript{2}H\textsubscript{6}  [\%] & 2.15 & 2.77 & 2.27 & 2.70 \\
Sel. of C\textsubscript{3}H\textsubscript{6}  [\%] & 6.32 & 8.00 & 8.65 & 9.93 \\
\hline
\end{tabular}
\caption{The conversion and selectivity of species in cases are considered in the CFD analysis. The flow rate of feed stream is fixed at $\dot{Q}^f= 10 \; \rm mL/min$. }
\label{tab:conversion_selectivity_of_two_inlet}
\end{center}
\end{table}

\subsubsection{Isolated Impact Analysis of Mixing and Heat Transfer in CFD}

\begin{figure}
   \centering
   \subfloat[ A (mixed, adiabatic wall) ]{ 
    \includegraphics[width=0.33\linewidth,trim={0cm 0cm 0cm 0cm},clip]{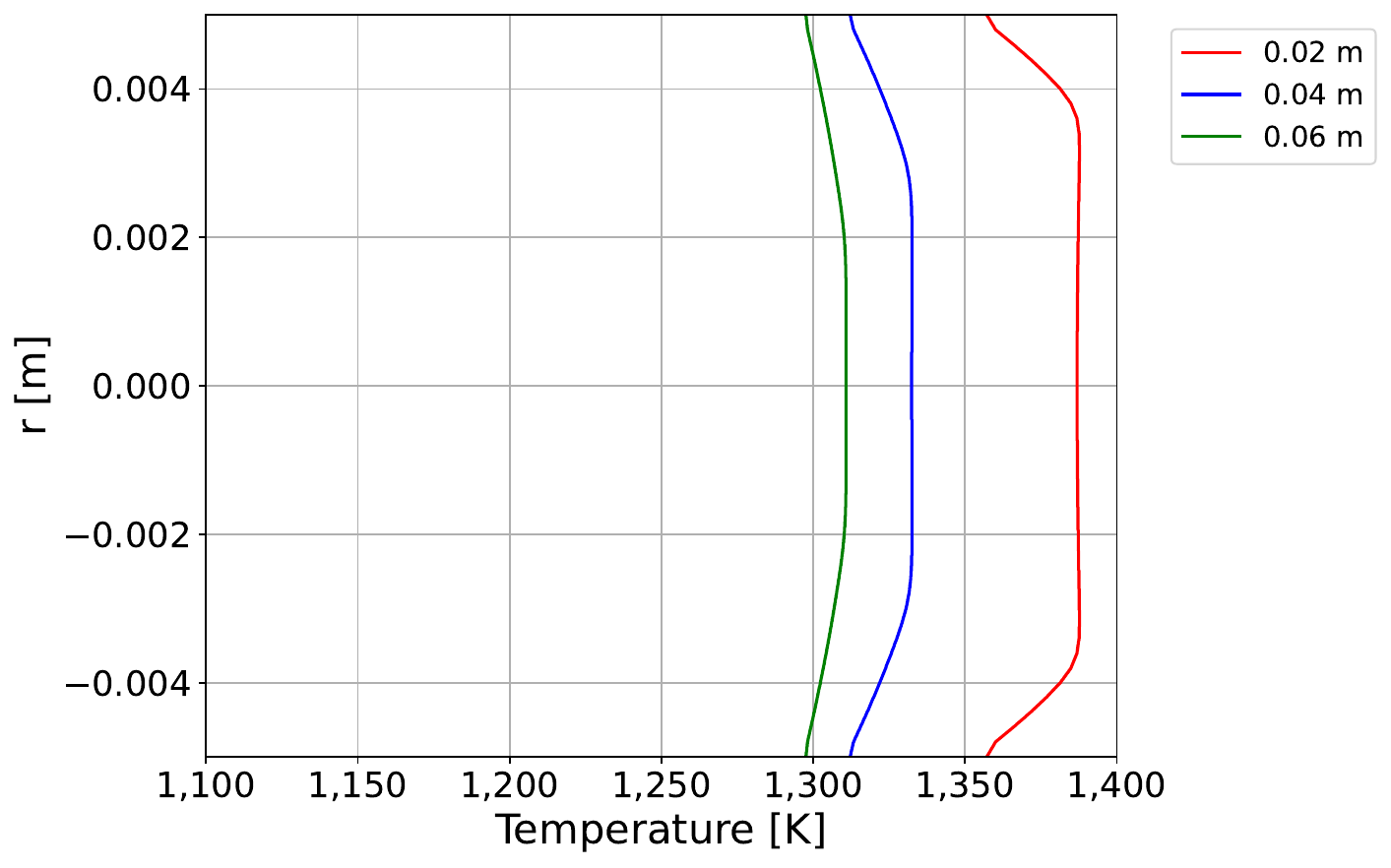}\label{subfig:temperatureprofiles_a}}
   \subfloat[  B  (mixed, a nonzero $h$) ]{
\includegraphics[width=0.33\linewidth,clip]{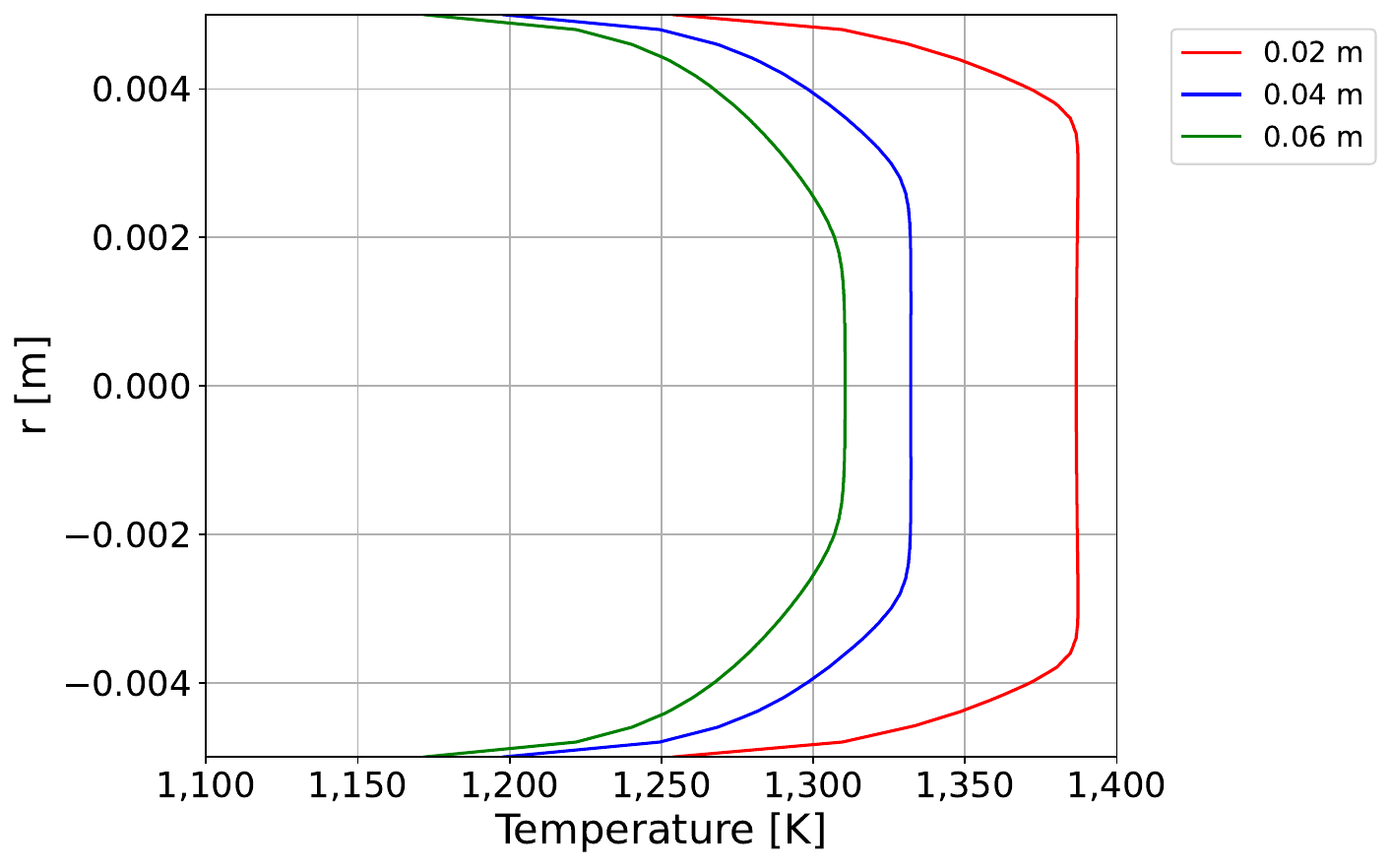}\label{subfig:temperatureprofiles_b}}
   \subfloat[ C (unmixed, adiabatic wall) ]{
\includegraphics[width=0.33\linewidth,clip]{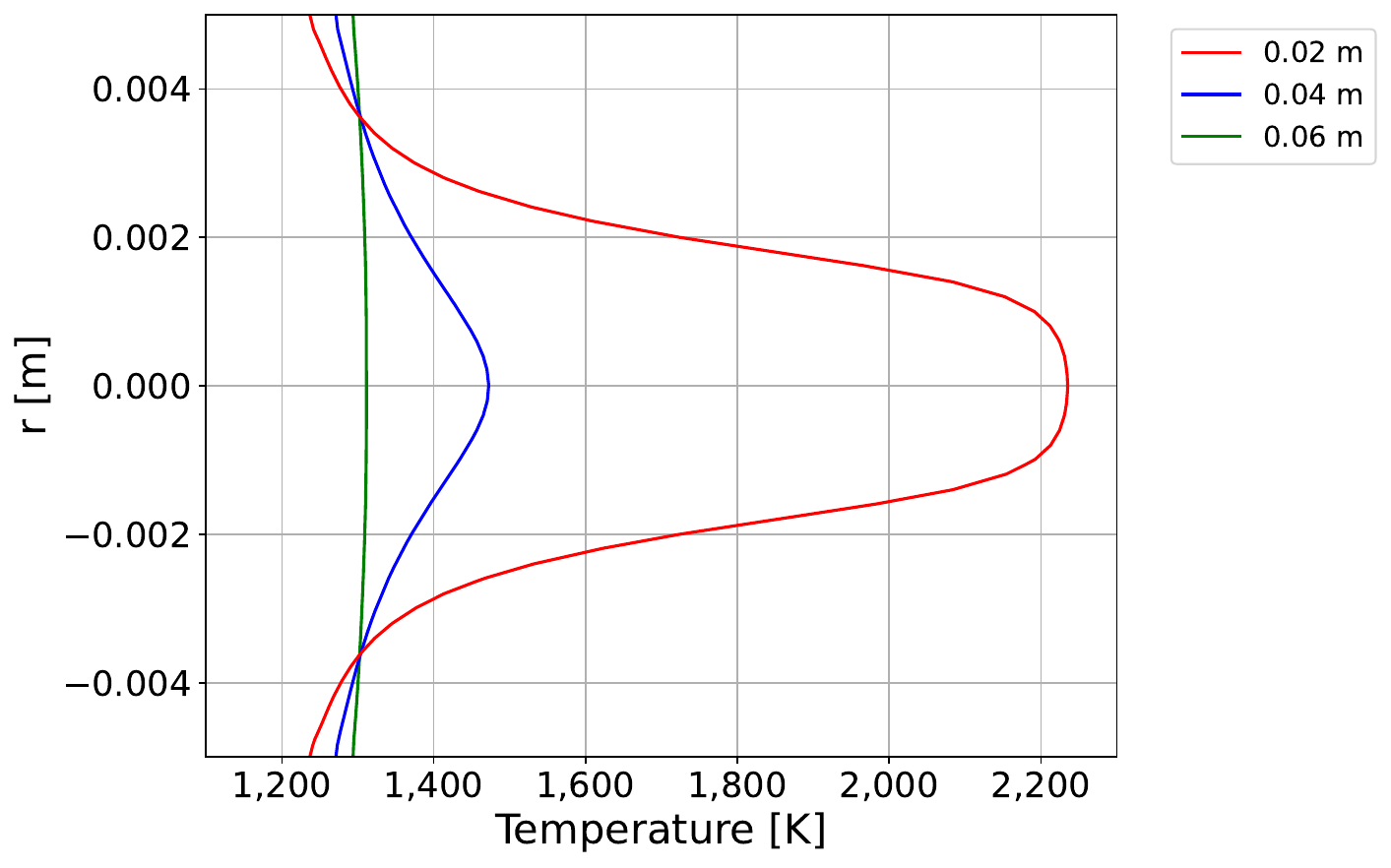}
   \label{subfig:temperatureprofiles_c}}
   \caption{Axial–radial temperature profiles in the early entrance region ($z < 0.06 \; \rm m$) for the cases listed in Table~\ref{tab:conversion_selectivity_of_two_inlet}.
   }
   \label{fig:temperatureprofiles}
\end{figure}

\begin{figure}
\begin{center}
\includegraphics[width=0.7\textwidth]{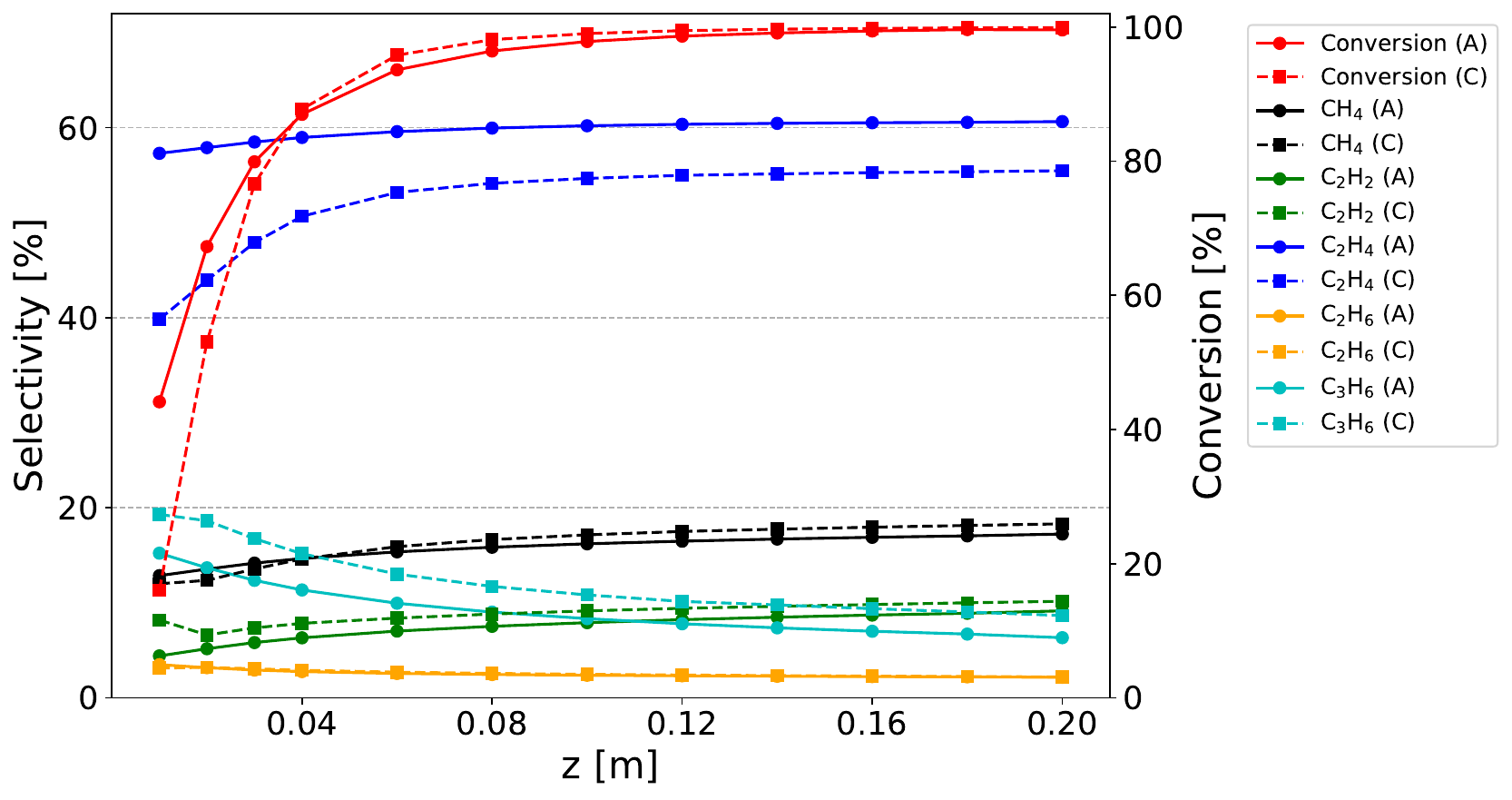}
\end{center}
\caption{Comparison of axial profiles of conversion and major product selectivity between CFD Cases A and C.}\label{fig:CFD_conversion_selectivity_AC}
\end{figure}

The discrepancy between the CFD and 1-D analyses is attributed to the spatially non-uniform temperature distribution in the radial direction, which the 1-D model does not capture.
In the CFD results, the feed gas flows along the reactor wall at a relatively low temperature (see \fref{fig:CFD_T_flow}), 
and the two gas streams do not reach thermal equilibrium 
at early entrance region, approximately  $z < 0.04 \; \rm m$.  
This results in steep radial temperature gradients near the inlet, causing the reaction to proceed at lower temperatures than those predicted by the 1-D analysis.

To better understand the separate impacts of heat loss through the wall and mixing condition, three additional simulations—representing intermediate cases closer to the assumptions of the 1-D model—are introduced.
\begin{itemize}
    \item Case A: premixed gas streams with adiabatic walls. 
    \item Case B: premixed gas streams and a heat transfer coefficient of $20\; \rm{W/(m^2\cdot{K})}$ for the surrounding wall $r=R$. 
    \item Case C: unmixed gas streams with adiabatic walls. 
    \item Full CFD: unmixed gas streams with a heat transfer.   
\end{itemize}
Note that Case A most closely resembles the 1-D analysis configuration, while the last scenario corresponds to the full CFD setup examined in the previous section.
In all cases, the feed flow rate $\dot{Q}^f$ is fixed at $10\; \rm{mL/min}$.
The CFD-predicted conversion and selectivity for the additional cases are summarized in Table~\ref{tab:conversion_selectivity_of_two_inlet}. 
To examine how mixing and heat loss influence the reactor's thermal behavior, we compare temperature profiles across all cases, as shown \fref{fig:temperatureprofiles}. 

A comparison between Cases A and B highlights the impact of heat loss to the environment on conversion and selectivity. As shown in \fref{subfig:temperatureprofiles_a} and \fref{subfig:temperatureprofiles_b}, Case B exhibits noticeable axial and radial heat loss, resulting in lower temperatures near the reactor wall compared to Case A. Although this temperature drop in the main pyrolysis region $(z<0.04\; \rm m)$ could influence reaction kinetics, Table~\ref{tab:conversion_selectivity_of_two_inlet} shows that the changes in conversion and selectivity are relatively minor—within 1-2~\%. 
This suggests that, under the given conditions, heat loss has a limited effect on product distribution.

\begin{figure}
\begin{center}
\includegraphics[width=0.7\textwidth]{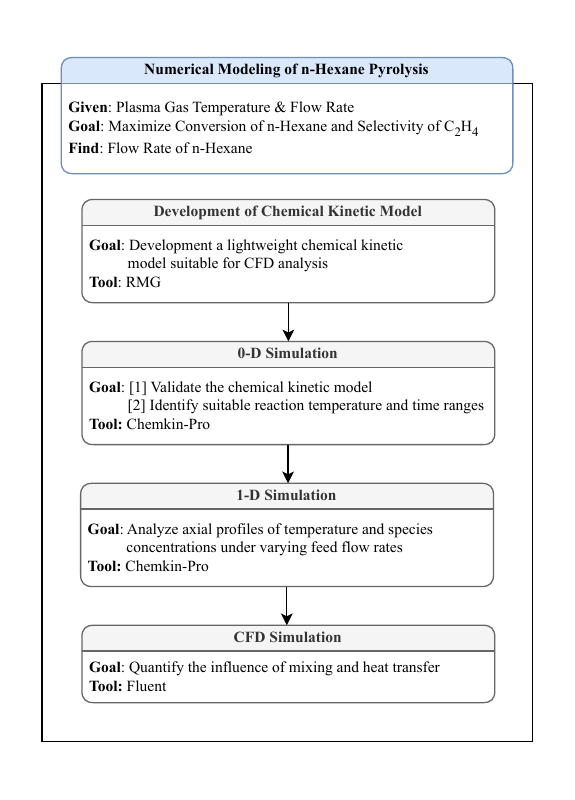}
\end{center}
\caption{Workflow chart of the simulation process to identify optimal operating conditions for maximizing $\rm{C}_2\rm{H}_4$ production efficiency relative to feed input in a plasma pyrolysis reactor.}\label{fig:workflowchart}
\end{figure}

On the other hand, the comparison between Cases A and C demonstrates that gas mixing significantly influences product selectivity.
As suggested in Table~\ref{tab:conversion_selectivity_of_two_inlet}, 
Case C exhibits reduced C\textsubscript{2}H\textsubscript{4} selectivity and increased C\textsubscript{3}H\textsubscript{6} selectivity.  
This shift in product distribution is attributed to differences in thermal conditions in the early entrance region ($z<0.04 \; \rm{m} $), where most \itn-hexane conversion occurs.
Pyrolysis is likely to occur along the boundary layers between the two unmixed gas streams, which are located near the reactor walls ($|r|>0.004\; \rm m$). 
In Case C, these regions exhibit lower temperatures (see  \fref{subfig:temperatureprofiles_a} and \fref{subfig:temperatureprofiles_c}), leading to different reaction kinetics.
According to the 0-D analysis results, such conditions favor C\textsubscript{3}H\textsubscript{6} formation.
Readers are referred to the cyan-colored line in \fref{fig:0d_result_comparison1}, which illustrates how these conditions promote C\textsubscript{3}H\textsubscript{6} production. 
The comparison of axial profiles of conversion and major product selectivities presented in \fref{fig:CFD_conversion_selectivity_AC} 
confirms this trend, showing elevated C\textsubscript{3}H\textsubscript{6} selectivity in the early entrance region for Case C. The results suggest that the mixing condition between the plasma discharge gas and the feed gas play a critical role in determining product selectivity.

In summary, the CFD analysis indicates that achieving high conversion requires careful control of the \itn-hexane flow rate. Moreover, optimizing product selectivity depends significantly on the extent of gas mixing. Therefore, a holistic approach that considers both flow rate control and mixing effects is essential for efficient pyrolysis.

\section{Conclusion}

In this study, the fundamental physics and chemical mechanisms governing a plasma pyrolysis reactor for \itn-hexane were systematically investigated.
The plasma environment was approximated as a high-temperature gas mixture of plasma discharge gases, allowing the complexities associated with detailed plasma chemistry to be circumvented.
Instead, emphasis was placed on the development of a tailored chemical kinetic mechanism for \itn-hexane pyrolysis, utilizing the RMG.
A lightweight chemical mechanism model was constructed to capture the essential phenomena of chemical reactions and was validated against well-established chemical mechanism sets. 

Recognizing the high computational cost associated with full CFD simulations across a broad design space, an incremental modeling strategy was adopted.
Initial 0-D and 1-D analyses were conducted to efficiently provide critical information on optimal flow rates for maximizing \itn-hexane conversion and target product selectivity.
Subsequently, CFD simulations incorporating heat transfer to the environment and gas mixing effects were performed, enabling more realistic predictions of reactor performance under plasma hydropyrolysis conditions.
The CFD results highlight that heat transfer and mixing conditions can substantially influence both the conversion and selectivity of the target products—effects that could not be captured by the 0-D and 1-D analyses alone.
Our approach
is summarized in \fref{fig:workflowchart}. 
This integrated approach
demonstrates that a combination of simplified and detailed models can accelerate the design optimization process and offers valuable insights for improving the pyrolysis efficiency of plasma reactors.

Building on these findings, future work will involve the experimental realization of reactor designs informed by the current modeling framework.
In particular, plasma reactors will be fabricated and tested based on the optimal conditions identified through this study.
Through iterative application of the developed design and analysis process—where modeling and experimental validation continuously inform each other—we aim to achieve a highly efficient plasma hydropyrolysis reactor.
This iterative design methodology is expected not only to refine reactor performance but also to establish a systematic framework for scaling up plasma-based hydropyrolysis technologies.

\section*{CRediT authorship contribution statement}

\textbf{Subin Choi:} Investigation, Formal analysis, Writing – original draft. 
\textbf{Chanmi Jung:} Conceptualization, Resources, Methodology
\textbf{Dae Hoon Lee:} Conceptualization, Project administration. \textbf{Jeongan Choi:} Supervision, Formal analysis, Conceptualization, Validation. 
\textbf{Jaekwang Kim:} Supervision, Formal analysis, Validation, Writing – review \& editing.

\section*{Declaration of competing interest}
The authors declare that they have no known competing financial
interests or personal relationships that could have appeared to influence
the work reported in this paper.

\section*{Acknowledgement}

This work was mainly supported by the Technology Innovation Program (or Industrial Strategic Technology Development Program-Carbon Neutral Industrial Core Technology Development Project) (RS-2023-00266831, Development of Novel Plasma Process for Hydrocarbon Cracking) funded By the Ministry of Trade, Industry \& Energy (MOTIE, Korea) and the National Research Foundation of Korea(NRF) grant funded by the Korea government(MSIT)(2022M3J8A1097255).
S. Choi was additionally supported by 2025 Hongik University Innovation Support Program Fund. J. Kim acknowledges support from the National Research Foundation of Korea grant funded by the Korean government under Grant No. RS-2024-00333943.

\appendix
\section{Appendix}

\subsection{The form for viscous tensor, the diffusive flux vector and the thermal flux vector}
\label{app:variousForms}

To construct a closed set of governing equations, 
we need to describe the viscous tensor $\bm{\tau}$, 
the diffusive flux vector $\mathbf{j}$, 
and the thermal flux vector $\mathbf{q}$
for mixture gases. 
In this work, the following forms are adopted from Ref~\cite{Bird}. Below, $X_i(= MM^{-1}_i Y_i$) 
 represents mole fraction of species $i$ in the mixture and $\mu_i$ represent the viscosity of $i$ species. 

\begin{itemize}
\item The viscous tensor $\bm{\tau}$, with the assumption of a Newtonian fluid is expressed as 
\begin{equation*}
    \bm{\tau} = \mu \left\{ (\nabla \bu + (\nabla \bu)^T)-\frac{2}{3}(\nabla \cdot \bm u ) \mathbf{I}
    \right\},
\end{equation*}
where the viscosity coefficient $\mu$ of a gas mixture is obtained by 
\begin{equation*}
    \mu =\sum^{N}_{i=1} X_i \mu_{i} \left(
    \sum^{N}_{j=i} X_j \phi_{ij}
    \right)^{-1}
\end{equation*}
where
\begin{equation*}
    \phi_{ij}=\left\{
\begin{aligned}
    &1  \quad\rm{if \;}i=j, \\
    &\frac{1}{\sqrt{8}}\left(1+\frac{M_i}{M_j} \right)^{-0.5} \left[
    1+\left( \frac{\mu_i}{\mu_j} \right)^{0.5} \left( \frac{M_j}{M_i}\right)^{0.25}
    \right]^2 \quad \rm othewise.
\end{aligned}
    \right.
\end{equation*}

\item The diffusive flux vector $\mathbf{j}$ of $i$-species is expressed generalizing the Fick's laws of diffusion  
\begin{equation*}
    \mathbf{j_i}=-\rho D_i \nabla  Y_i
\end{equation*}
where the diffusion coefficients $D_i$ of all species are written in terms of the binary diffusion coefficient matrix
\begin{equation*}
    D_i =(1-Y_i)^{-1} \left( \sum_{i \neq j} D^{-1}_{ij}X_i
    \right)^{-1}.
\end{equation*}

\item The thermal flux vector $\mathbf{q}$ is expressed as the combined effect of the conduction and mass diffusion
\begin{equation*}
    \mathbf{q} = -\lambda \nabla T + \sum^{N}_i h_i \mathbf{j}_i,
\end{equation*}
where the coefficient of heat conductivity $\lambda$ of the mixture is obtained through a combination averaging formula~\cite{Bird}
\begin{equation*}
    \lambda = \frac{1}{2} \left( \sum^{N}_{i=1} X_i \lambda_i + \frac{1}{\sum^{N}_{i=1} X_i/\lambda_i} \right).
\end{equation*}

\end{itemize}

\subsection{The net production rate of each chemical species}
\label{app:productionRate}

Here, we provide the form for computing the net mass production rate $\dot{R}_i$ predicted by a set of stoichiometric chemical reactions. 
For illustration, consider a chemical reaction of reactant $A$ and $B$ that producing $C$ and $D$, 
\begin{equation*}
aA + bB \to cC + dD,   
\end{equation*}
where $a,b,c$ and $d$ are stoichiometric coefficients of each species. 
A general form elementary chemical reactions are described by 
\begin{equation}
    \sum^{N}_i \nu'_i m_i \underset{k_f}{\stackrel{k_b}{\rightleftharpoons}} \sum^{N}_i \nu''_i m_i
\end{equation}
where $\nu'_i$ and $\nu''_i$ are stoichiometric coefficients for reactants and products respectively. 
$k_f$ stands for the reaction rate of the forward reaction, while $k_b$ means a reaction rate for the backward direction. 
Then, the net production of a given $i$-species, or $R_i$, is computed through 
\begin{equation*}
    R_i = M_i \sum^{N_r}_{k=1}(\nu''_{i,k}- \nu'_{i,k}) \dot R_k
\end{equation*}
where $M_i$ denotes the molecular mass of species, $N_r$ means the total number of elementary chemical reaction stages. The symbol $\dot R_k$ denotes the progress rate of $k$-th stage of the elementary chemical reaction, which is evaluated as 
\begin{equation}
    \dot{R} _k = k_{f,k}\prod^{N}_i C^{\nu'_{i,k}}_i - k_{b,k}\prod^{N}_i C^{\nu''_{i,k}}_i,  
\end{equation}
where $k_{f,k}$ (or $k_{b,k}$) represents the reaction rate coefficient for forward (or backward) reaction of the $k$-th stage of the elementary reaction, and $C_i=Y_i\rho M_i$ is the concentration of species $i$.
The reaction rate coefficients are given by the empirical Arrhenius-law. For example, $k_{f,k}$ is in the form of 
\begin{equation}
    k_{f,k}= B_{f,k} T^{a_f,k}\exp\left( 
    -\frac{E_{f,k}}{RT} 
    \right),
    \label{eqn:kfk}
\end{equation}
where $B_{f,k}$ is frequency factors, $\alpha_{f,k}$ is  temperature indices, and $E_{f,k}$ is the activation energy of the forward reaction.
Reaction data should include these constants.
The form of $k_{b,k}$ has the same structure with Eq.~\eref{eqn:kfk}, with $f$ replaced by $b$.

\printbibliography

\end{document}